\documentclass[lettersize,journal]{IEEEtran}

\ifCLASSOPTIONcompsoc
\usepackage[nocompress]{cite}
\else
\usepackage{cite}
\fi
\ifCLASSINFOpdf
\else
\fi
\usepackage{stmaryrd}
\usepackage{pifont}
\usepackage{cite}
\usepackage{bbding}
\usepackage{bm}
\usepackage{multirow}
\usepackage{amssymb}
\usepackage{amssymb}
\usepackage{balance}
\usepackage{amsmath}
\usepackage{amsfonts}
\usepackage{amssymb}
\usepackage{algorithm}
\usepackage{multirow}
\usepackage{algpseudocode}
\usepackage{latexsym}
\usepackage{graphicx}
\usepackage{epsfig}
\usepackage{float}
\usepackage{booktabs}
\usepackage{flushend}
\usepackage{array}
\usepackage{color}
\usepackage{caption,subcaption}
\usepackage{graphicx}
\usepackage[english]{babel}
\usepackage[table]{xcolor}
\usepackage[pagebackref=true,breaklinks=true,letterpaper=true,colorlinks,bookmarks=false]{hyperref}
\usepackage{float}
\usepackage{epstopdf}

\usepackage{listings}
\makeatletter
\newcommand*\bigcdot{\mathpalette\bigcdot@{.5}}
\newcommand*\bigcdot@[2]{\mathbin{\vcenter{\hbox{\scalebox{#2}{$\m@th#1\bullet$}}}}}
\makeatother

\definecolor{c2}{HTML}{FBD9BD}

\begin{document}
	%
	% Arbitrary
	\title{HQG-Net: Unpaired Medical Image Enhancement with High-Quality Guidance}
	%\title{Guiding with Explicit High-quality Cues: A Variational Guidance Perspective for Unpaired Medical Image Enhancement}
	%\title{XXGAN: Unpaired Medical Image Enhancement with Explicit High-quality Guidance}
	%\title{Explicit High-quality Guidance is a Good Assistant for Robust Unpaired Medical Image Enhancement}
	%\title{Towards Robust and Adaptive Unpaired Medical Image Enhancement: A High-quality Variational Guidance Perspective}
	%\title{How to Exploit the Unpaired High-quality Information? Explicit Variational Guidance for Unpaired Medical Image Enhancement}
	% \author{Chunming He\inst{1} \and
		% Kai Li\inst{2} \and
		% Guoxia Xu \inst{3} \and
		% Jiangpeng Yan \inst{1} \and
		% Longxiang Tang\inst{4} \and
		% Yulun Zhang\inst{5} \and
		% Xiu Li\inst{1}}
	% %
	% \authorrunning{C. He et al.}
	% \institute{Tsinghua Shenzhen International Graduate School, Tsinghua University, China  \\\and
		% NEC Laboratories America, USA \and
		% Norwegian University of Science and Technology, Norway \and
		% University of Electronic Science and Technology of China, China  \and
		% ETH Zurich, Switzerland
		%  }

%

\author{
	Chunming He,~\IEEEmembership{}
	Kai Li,~\IEEEmembership{Member,~IEEE,}
	Guoxia Xu,~\IEEEmembership{Member,~IEEE,}
	Jiangpeng Yan,~\IEEEmembership{}
	Longxiang Tang,~\IEEEmembership{}
	Yulun Zhang,~\IEEEmembership{Member,~IEEE,}
	Xiu Li,~\IEEEmembership{Member, ~IEEE,}
	and Yaowei Wang,~\IEEEmembership{Member, ~IEEE}
	% <-this % stops a space
	\thanks{This work was supported by NSFC 41876098 and Shenzhen Science and Technology Project. (Grant No. JCYJ20200109143041798, WDZC20200820200655001). \textit{(Corresponding author: Xiu Li) (E-mail: li.xiu@sz.tsinghua.edu.cn)}}
	\thanks{Chunming He, Jiangpeng Yan, Longxiang Tang, and Xiu Li are with Tsinghua Shenzhen International Graduate School, Tsinghua University, Shenzhen 518055, China. Kai Li is with Machine Learning Department, NEC Laboratories America Inc., NJ 08540, America. Guoxia Xu is with Department of Computer Science, Norwegian University of Science and Technology, 2815 Gjovik, Norway. Yulun Zhang is with the Computer Vision Lab, ETH Z{\"u}rich, Z{\"u}rich 8092, Switzerland. Yaowei Wang is with Peng Cheng Laboratory, Shenzhen 518066, China.}
}
	\maketitle              % typeset the header of the contribution
	%
	
	%\IEEEtitleabstractindextext{%
\begin{abstract}

Unpaired Medical Image Enhancement (UMIE) aims to transform a low-quality (LQ) medical image into a high-quality (HQ) one without relying on paired images for training. While most existing approaches are based on Pix2Pix/CycleGAN and are effective to some extent, they fail to explicitly use HQ information to guide the enhancement process, which can lead to undesired artifacts and structural distortions. In this paper, we propose a novel UMIE approach that avoids the above limitation of existing methods by directly encoding HQ cues into the LQ enhancement process in a variational fashion and thus model the UMIE task under the joint distribution between the LQ and HQ domains. Specifically, we extract features from an HQ image and explicitly insert the features, which are expected to encode HQ cues, into the  enhancement network to guide the LQ enhancement with the variational normalization module. We train the enhancement network adversarially with a discriminator to ensure the generated HQ image falls into the HQ domain. We further propose a content-aware loss to guide the enhancement process with wavelet-based pixel-level and multi-encoder-based feature-level constraints. Additionally, as a key motivation for performing image enhancement is to make the enhanced images serve better for downstream tasks, we propose a bi-level learning scheme to optimize the UMIE task and downstream tasks cooperatively, helping generate HQ images both visually appealing and favorable for downstream tasks. Experiments on three medical datasets, including two newly collected datasets, verify that the proposed method outperforms existing techniques in terms of both enhancement quality and downstream task performance. We will make the code and the newly collected datasets publicly available for community study.
	\end{abstract}
	

	% is not selected <OR> if conference mode is selected - because compsoc
	% conference papers position the abstract like regular (non-compsoc)
%	\IEEEdisplaynontitleabstractindextext

	% \fi
	\IEEEpeerreviewmaketitle

	%\ifCLASSOPTIONcompsoc
	%\IEEEraisesectionheading{\section{Introduction}\label{sec:introduction}}
	%\else
	%\section{Introduction}
	%\label{sec:introduction}
	%\fi
    %% narrow the gap between equations and sentences
    \setlength{\abovedisplayskip}{2pt}
    \setlength{\belowdisplayskip}{2pt}
	\section{Introduction}\label{sec:introduction}
\begin{figure}[htbp!]
 		\centering
 		\includegraphics[width=0.45\textwidth]{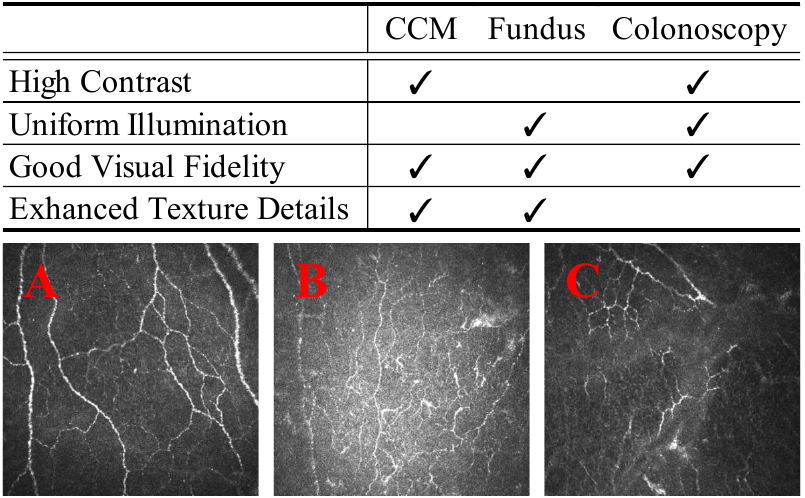}
 		\vspace{-2mm}
 		\caption{HQ definitions on different medical data~\cite{ma2021structure,shen2021hrenet,zhou2021study}. For instance, in Corneal Confocal Microscopy (CCM) data, image A is considered HQ, while B and C are the LQ images. B suffers from low contrast and C has blurred texture details. Despite the degeneracy, all three images still exhibit intra-modality homogeneity.}
 		\label{fig:HQDefinition}
 		\vspace{-0.5cm}
 	\end{figure}

	\IEEEPARstart{D}~ue to the variability of light transmission and clinical imaging conditions, medical images often exhibit uneven illumination or blurry texture details~\cite{ma2021structure,ikuta2022deep,xu2023dm} (see Fig.~\ref{fig:HQDefinition} B and C). These low-quality (LQ) images can significantly impede automated disease screening, examination, and diagnosis. Medical image enhancement aims to
	transform an LQ image to a high-quality (HQ) one that fulfills the modality-specific HQ definitions (in Fig.~\ref{fig:HQDefinition}), such as enhanced illumination quality and texture details~\cite{ji2021progressively,mou2019cs,he2022learning}. Early approaches optimize energy-based objective functions based on certain imaging priors, such as Retinex equation~\cite{jeelani2019content,zhao2020automated} and region line prior~\cite{ju2021idrlp}. However, these methods are limited in their ability to generalize to different types of degradation conditions, as one imaging prior may only apply to certain types of degradation conditions, restricting their applicability.
    % one imaging prior usually applies to only certain types of degradation conditions and cannot generalize to others, which limits the application scope of these approaches. 
    Recently, deep learning-based methods provide new avenues to this problem and use neural networks to approximate various degradation models. A naive approach involves training a neural network, such as Pix2Pix~\cite{isola2017image}, to translate an LQ image from the LQ domain to the HQ domain at the pixel level. However, this approach requires paired LQ-HQ images with pixel-to-pixel correspondence for training, which can be challenging to obtain for medical applications \cite{ma2021structure,he2023camouflaged,he2023weaklysupervised,XU2022102643}.
	
	\begin{figure*}[htbp!]
		\centering
		\includegraphics[width=\textwidth]{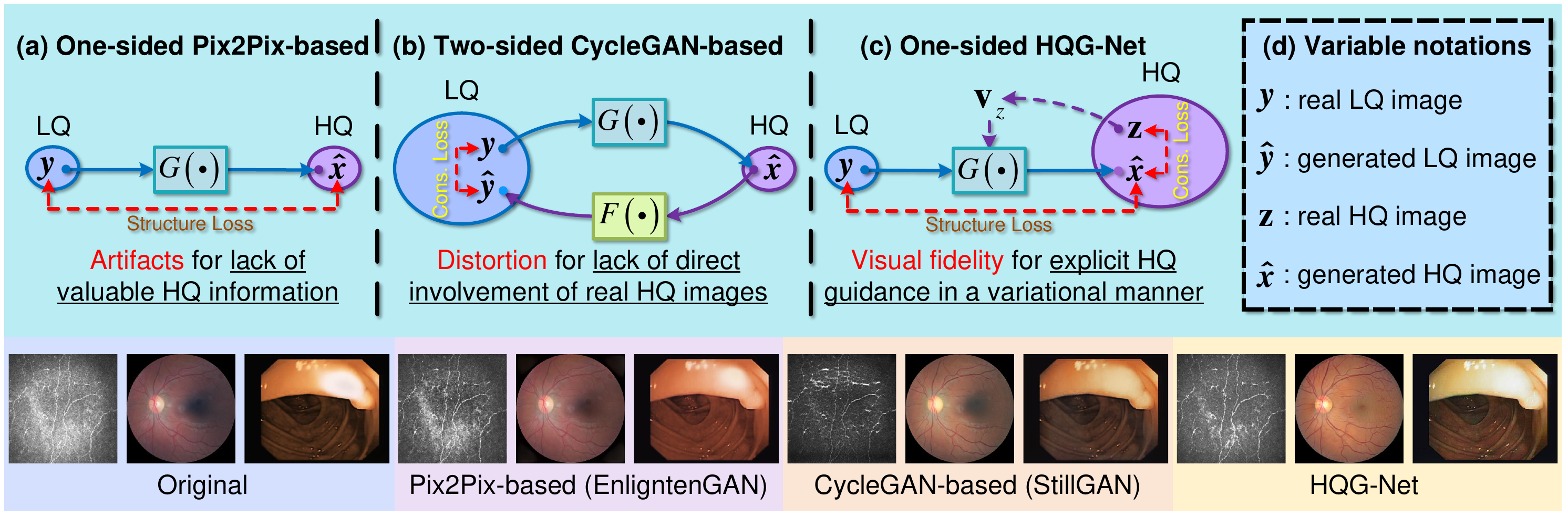}
		\vspace{-7mm}
		\caption{Comparison of different categories of UMIE methods. (a) Pix2Pix-based methods learn a one-sided mapping from the LQ domain to the HQ domain guided by some structure losses. These methods fail to exploit the valuable information from unpaired HQ images. (b) CycleGAN-based approaches are proposed to learn a bi-directional pixel-level map between LQ and HQ domains. (c) the proposed HOG-Net introduces the explicit HQ guidance by injecting the feature vector $\mathbf{v}_z$ of an unpaired HQ sample $\mathbf{z}$ into the enhancement generator $G\left(\bigcdot\right)$ with two task-specific loss functions.
		}
		\vspace{-0.5cm}
		\label{fig:Comparison}
	\end{figure*}

	To address the enhancement task more practically, several methods have investigated the unpaired medical image enhancement (UMIE) problem, which requires only unpaired HQ and LQ images as training data. The most intuitive idea is to retain the main structure of Pix2Pix~\cite{isola2017image} but use specific loss functions for inter-domain structure preservation, e.g., structure loss (Fig.~\ref{fig:Comparison} (a)). For instance, EnlightenGAN~\cite{jiang2021enlightengan} employs a structure-preserving loss with a global-local discriminator. However, the generator of the Pix2Pix-based method is mainly trained with an unsupervised loss between real LQ and generated HQ images, which neglects valuable real HQ information during the training phase, leading to the generation of artifacts in the enhanced medical images. Another approach is to exploit CycleGAN~\cite{zhu2017unpaired}, which involves a two-sided translation between LQ and HQ domains for the intra-domain distribution similarity with a cycle-consistent constraint (Fig.~\ref{fig:Comparison} (b)). By employing both HQ and LQ images for network training, CycleGAN-based techniques outperform Pix2Pix-based methods in exploiting valuable information from unpaired real HQ images. However, since the real HQ images are not directly involved in the enhancement of LQ images, there exists a domain shift between the enhanced LQ (generated HQ) and the real HQ domains, leading to potential influences on structure fidelity and texture distortion. For structure preservation, StillGAN~\cite{ma2021structure}, a CycleGAN-based approach, proposes an SSIM-based structure loss. However, this loss function is applied to the entire image space and can hardly distinguish the complete structure information of LQ medical images with complex degradation.

\begin{figure*}
		\centering
		\includegraphics[width=1\textwidth]{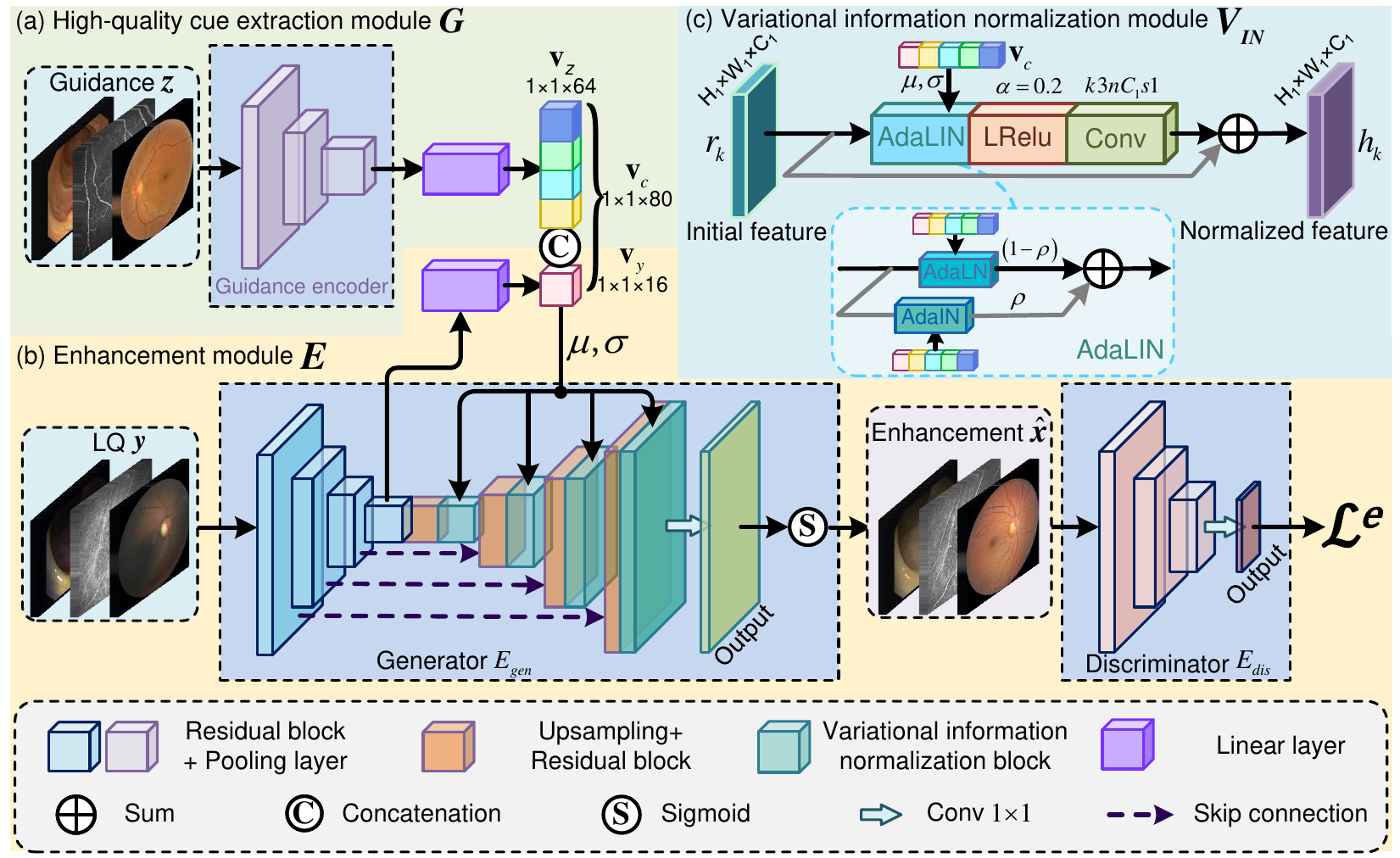}\vspace{-0.2cm}
		\caption{Framework of the proposed HQG-Net. ``LRelu'' and ``Conv" are short for LeakyReLU and convolution, where $k$, $n$, $s$ in $k3nC_1s1$ denote kernel, channel, stride. For clarity, we omit other inputs to the discriminator. Following \cite{choi2020stargan} and~\cite{ma2021structure}, the backbone of the high-quality cue extraction module $G$ and the generator of the enhancement module $E_{gen}$ are ResNet-18 and ResUNet50. The encoder and decoder of $E_{gen}$ are denoted as $E_e$ and $E_d$, where $E_d$ has several extra variational information normalization (VIN) modules to inject HQ cues. In $E_{gen}$, feature maps $f_k$, $r_k$, $h_k$ ($k\in\{1,2,3,4\}$) are generated by the basic encoder (the blue blocks), the basic decoder (the orange blocks), the VIN modules (the dark green blocks), respectively.}
		\label{fig:Framework}
		\vspace{-3mm}
	\end{figure*}

    Considering the intra-modality homogeneity of medical images (Fig.~\ref{fig:HQDefinition}), it is desirable to enhance LQ images with the guidance of unpaired but homogeneous HQ images. Therefore, we propose HQG-Net, a GAN-based UMIE network that explicitly utilizes unpaired HQ images as guidance for LQ image enhancement (see Fig.~\ref{fig:Comparison} (c)). To ensure the stability of HQG-Net, we introduce the feature vector $\mathbf{v}_z$ of an unpaired HQ sample $\mathbf{z}$ as guidance. This feature vector is a condensed extraction of HQ information and can moderate the slight degradation from the HQ sample. To preserve structure information from the LQ input, we propose a Variational Information Normalization (VIN) module (see Fig.~\ref{fig:Framework} (c)) that guides the enhancement in a variational manner, where the concise and general variational information ensures the structure preservation of the original image during the domain translation. By introducing the variational guidance information, HQG-Net models the UMIE task under a joint distribution between the LQ and HQ domains, which is supposed to learn more comprehensive and unbiased information~\cite{yue2020dual,Xu_Wang_Nie_Li_2023} and thereby ensures visual fidelity~\cite{xie2022joint}.

	Furthermore, we propose a content-aware loss to enhance texture details and improve visual fidelity for automated disease screening and diagnosis, incorporating both pixel-level and feature-level constraints. Specifically, the wavelet-based pixel-level constraint ensures inter-domain structure preservation in the high-frequency component, which contains the most abundant texture information, rather than in the whole image space, encouraging the network to focus on translating critical structures from the LQ source image. To address the limited information representation of the pixel-level cycle-consistent loss, we propose a multi-encoder-based feature-level regularization is proposed for intra-domain distribution similarity with deep feature consistency.
    Due to the domain gap between low-level and high-level vision tasks, a high-quality reconstruction result may not  necessarily guarantee good performance on downstream tasks~\cite{yu2022healthnet,fernandes2020automatic} (see Fig.~\ref{fig:CCM_Quanti}). To alleviate this, we propose a cooperative training strategy with a bi-level learning scheme that jointly learns the UMIE task and the downstream tasks such as medical image segmentation and medical image classification. The aim is to generate HQ enhanced results that are both visually appealing and favorable for the downstream tasks.
	
	To comprehensively evaluate the proposed method, we collect two datasets (named \textit{Fundus} and \textit{Colonoscopy} datasets) in addition to the one currently used in this field~\cite{ma2021structure}. Our experiments on the three datasets comprehensively show that HQG-Net consistently outperforms the existing methods. 
	
Our contributions are summarized in the following respects: 
\begin{itemize}
    \item We propose HQG-Net, which incorporates HQ cues to guide LQ image enhancement. In this way, we model UMIE under the joint distribution between LQ and HQ domains that contains more complete and unbiased information, and thus ensures the stability of the framework and improves the visual fidelity of the enhanced result. 
\item We introduce a content-aware loss for textural enhancement and visual fidelity with the wavelet-based pixel-level and the multi-encoder-based feature-level constraints.

\item We propose a plug-and-play cooperative training strategy with a bi-level optimization formulation to produce a better downstream performance as well as enhanced results with better visually appealing effects. 

\item We conduct extensive experiments on three datasets, including two newly collected datasets, and the results show that HQG-Net achieves state-of-the-art (SOTA) performance in terms of enhancement quality, downstream task performance, and computational efficiency. The newly collected datasets will be made publicly available.

\end{itemize}

\section{Related Work}\label{sec:relatedwork}%\vspace{-1mm}

\subsection{Medical Image Enhancement}
Approaches to medical image enhancement can be broadly categorized into model-based and learning-based methods. Model-based methods involve exploiting a specific manually designed imaging prior and then empirically correcting the LQ image correction. Early techniques focused on the global enhancement of the LQ images through stretching the parameters of dynamic ranges, such as histogram equalization (HE)~\cite{abdullah2007dynamic}, image contrast normalization (ICN)~\cite{foracchia2005luminosity}, and contrast limited adaptive histogram equalization (CLAHE)~\cite{setiawan2013color}. Retinex-based methods~\cite{jeelani2019content,zhao2020automated} were later proposed by decomposing the LQ image into reflectance and illumination parts and recovering them with the corresponding prior knowledge. Additionally, IDRLP~\cite{ju2021idrlp} discovered a region line prior to enhance the LQ images with a region fidelity constraint. However, those traditional methods are primarily based on  manually designed imaging priors, which fail to handle the complex degradation conditions and thus suffer from the limited application scope.

Learning-based methods are more flexible and can be extended to the unpaired setting, where paired LQ-HQ images are not required for model training. Intuitively, the unpaired problem can be solved by the structure of Pix2Pix~\cite{isola2017image} with task-specific loss functions in an unsupervised framework~\cite{jiang2021enlightengan,liao2019adn}. ADN~\cite{liao2019adn} employed multiple encoders and decoders to separate the content information from the degraded component and train the network with specialized loss in an unsupervised manner. EnlightenGAN~\cite{jiang2021enlightengan} proposed a multi-scale discriminator with a structure-preserving loss. Nevertheless, those Pix2Pix-based networks neglect the valuable information from the unpaired HQ images, which weakens their resilience to complex degradation conditions. To exploit the unpaired LQ and HQ images, CycleGAN~\cite{zhu2017unpaired} was proposed with the cycle consistency constraint to learn a mapping between the LQ and the HQ domains. StillGAN~\cite{ma2021structure} was a CycleGAN-based method with a structure preservation loss and an illumination constraint. Unfortunately, the two-sided cycle-consistency strategy lacks the direct involvement of HQ images in the enhancement of LQ images, which inevitably brings domain shift and thus results in structure distortion~\cite{men2022unpaired}. {To solve these problems, we propose the explicit guidance-based HQG-Net with the variational information normalization module and the content-aware loss for structure preservation and visual fidelity. Furthermore, we design a plug-and-play cooperative training strategy to simultaneously acquire visual-appealing enhancement results and satisfactory downstream applications.}

\subsection{Reference-based Super-Resolution and Style Transfer} 
Our HQG-Net is also related to reference-based image super-resolution (Ref-SR) and Style Transfer (Ref-ST), which mainly focus on natural images. While it may seem straightforward to extend methods from these fields to medical images, we argue that such naive extension often fails to produce satisfactory results as the unique characteristics of medical images are not properly modeled. In reference-based image super-resolution (Ref-SR), the generation of texture details heavily relies on the reference image via point-wise or patch-wise correspondence. CrossNet~\cite{zheng2018crossnet} established a relationship between the reference image and the low-resolution (LR) image by flow estimation for point-wise alignment. For patch matching, NTT~\cite{zhang2019image} matched the multi-level features extracted from a pretrained VGG between the reference image and LR image in a concatenated manner. Style transfer aims to generate a re-stylized image by rendering the source image with the style component from a given reference image. By embedding the encoded reference image into a latent space, StyleGAN~\cite{karras2019style} enabled controlled modifications in style transfer. To generate diverse images across multiple domains, StarGAN V2~\cite{choi2020stargan} introduced a style mapping network with domain-specific style codes extracted by specific reference images. 

It should be noted that we are the \textbf{first} to handle the UMIE problem with the explicit assistance of reference images. Besides this, HQG-Net differs from existing Ref-SR and Ref-ST methods in several aspects. \textbf{First}, unlike source images in ref-SR or ref-ST that are of high quality, LQ images in UMIE often suffer from complex degradation conditions~\cite{shen2020modeling}, such as light transmission disturbance and environment-induced artifacts. {Therefore, a stable HQ cue extractor and general guidance are required to handle the complex degradation. To achieve this, we pretrain the HQ cue extraction module $G$ with images from both LQ and HQ domains to achieve stable HQ cue extraction and guide the enhancement in a condensed variational manner with the purpose of general guidance.} \textbf{Second}, different from ref-ST and ref-SR, which prioritize visually appealing results, UMIE also emphasizes structure preservation for clinical requirements~\cite{ma2021structure}. However, both the feature-level alignment (ref-SR) and the reference-only guidance manner (ref-ST) can inevitably suppress the structure preservation capacity. {To accommodate this requirement, we propose the task-specific content-aware loss and make two additional changes to our HQG-Net, including guiding the enhancement with the joint variational vector $\mathbf{v}_c$ (including $\mathbf{v}_z$ and $\mathbf{v}_y$), and using the adaptive normalization operator $AdaLIN$.} \begin{table}[htbp!]
\centering
\caption{Summation of the primary notations used in this paper.}
\label{table:notation}
\resizebox{\columnwidth}{!}{
\setlength{\tabcolsep}{1mm}
\begin{tabular}{ll} \toprule[1.5pt]
Notations                                      & Meanings                                                            \\ \hline \hline
\multicolumn{2}{l}{\textbf{Data}}                     \\ \hline
$\mathbf{z}$                                   & HQ guidance image                                                   \\
$\mathbf{y}$                                   & LQ image                                                            \\
$\widehat{\mathbf{x}}$                         & Enhanced result of the LQ image                                     \\ \hline
%$\widehat{\mathbf{x}}^{HF}$, $\mathbf{y}^{HF}$ & High-frequency component of $\widehat{\mathbf{x}}$ and $\mathbf{y}$ \\ \hline
\multicolumn{2}{l}{\textbf{Guidance vectors}}               \\ \hline
$\textbf{v}_z$                                 & HQ guidance vector                                                  \\
$\textbf{v}_y$                                 & LQ structure vector                                                 \\
$\textbf{v}_c$                                 & Joint vector                                                        \\ \hline
\multicolumn{2}{l}{\textbf{Feature maps}}            \\ \hline 
$f_k$     & Feature maps of the encoder $E_e$  \\
\multirow{2}{*}{$r_k$, $h_k$}                  & Feature maps of the decoder $E_d$ before and after being        \\
 & processed by the variational information normalization module  \\ \bottomrule[1.5pt]
\end{tabular}}\vspace{-3mm}
\end{table}{By employing the joint vector and the adaptive normalization operator, we expect the enhanced results to simultaneously ensure high-quality reconstruction performance and preserve the critical structure information with the joint guidance strategy and the learnable hybrid normalization operation.}

\section{Methodology}%\vspace{-1mm}
Unpaired Medical Image Enhancement (UMIE) aims to learn an enhancement model with a low-quality (LQ) image dataset $\mathcal{Y}$ and a high-quality (HQ) image dataset $\mathcal{Z}$. Note that $\mathcal{Y}$ and $\mathcal{Z}$ are unpaired, i.e., there is not an image from $\mathcal{Z}$ that is the high-quality version of an image from $\mathcal{Y}$. We propose a novel UMIE technique, dubbed HQG-Net, by explicitly employing HQ information to guide the enhancement of the LQ image. For structure preservation and visual fidelity, a novel content-aware loss is designed with the wavelet-based pixel-level constraint and the multi-encoder-based feature-level regularization. Furthermore, we propose a cooperative training strategy with a bi-level optimization formulation to produce a better downstream performance as well as visually appealing enhanced results. {For better comprehension, we summarize the primary notations utilized in this paper, along with their corresponding meanings, in Table~\ref{table:notation}.}

\subsection{Network Architecture}
HQG-Net consists of four main parts, including a high-quality cue extraction module $G$, a generator $E_{gen}$ of the enhancement module $E$, a variational information normalization module $V_{IN}$, and {a discriminator $E_{dis}$}, where $V_{IN}$ convey the HQ cue from $G$ to $E$ for high-quality (HQ) guidance. The specific network architecture of HQG-Net is shown in Fig. \ref{fig:Framework}.
	
\noindent\textbf{High-quality cue extraction}. 
An HQ image $\mathbf{z}$ is randomly selected and then feed into $G$ to extract high quality cues,
	\begin{equation}
	    \textbf{v}_z = G(\mathbf{z}), 
	\end{equation}
	where $\mathbf{z}\in\mathcal{Z}$, $\textbf{v}_z\in\mathbb{R}^{d_z}$ is the vector with h-dimension expected to encode HQ cue and $h=64$. Following \cite{choi2020stargan}, we implement $G$ as ResNet-18, plus a full-connected (FC) layer to get the guidance vector. To facilitate HQ cue extraction, we pretrain ResNet-18 within an auto-encoder framework. Specifically, we use ResNet-18 as an encoder, append a decoder with the transposed structure, and pretrain the encoder-decoder model with the standard MSE loss and SSIM loss~\cite{liu2022target}. {Apart from the HQ images from $\mathcal{Z}$, we pretrain the model with additional LQ images from $\mathcal{Y}$ to ensure its robustness to the latent subtle perturbations in the HQ guidance image and thus facilitate the extraction of stable HQ cues.}

\noindent\textbf{Guided low-quality image enhancement}. 
{Following~\cite{ma2021structure}, our enhancement network is based on ResUNet50 with the last layer abandoned for computational efficiency. Besides, we plus several extra Variational Information Normalization (VIN) Modules that inject HQ cues in multiple levels. Specifically,  given an LQ image $\mathbf{y}\in\mathcal{Y}$, the ResNet50 encoder $E_e$ generates a set of feature maps $f_k$ ($k\in \left\{1,2,3,4\right\}$) in different levels.}

Considering that the latent domain discrepancy between the LQ input $\mathbf{y}$ and the HQ guidance $\mathbf{z}$ can affect the quality of image enhancement, it is desirable to generate the enhanced result $\widehat{\mathbf{x}}$ with the joint guidance of $\mathbf{y}$ and $\mathbf{z}$ to simultaneously ensure the high-quality reconstruction and preserve the significant semantic information. {Therefore, we map $f_4$ to a compact vector $\textbf{v}_y\in\mathbb{R}^{16}$, which is expected to contain critical structural information, by a linear layer $FC_e$, i.e., $\textbf{v}_y = FC_e\left(E_{e}(\mathbf{y})\right)$.} We concatenate $\textbf{v}_y$ with the extracted HQ vector $\textbf{v}_z$ and thus obtain the integrated joint vector $\textbf{v}_c$
\begin{equation}
	 \textbf{v}_c = concate\left(\textbf{v}_z,\textbf{v}_y\right),
\end{equation}
which encodes the gist information from both the LQ image $\mathbf{y}$ and the HQ image $\mathbf{z}$, and we use $\textbf{v}_c$ as one input for the VIN modules inserted into the decoder in a multi-scale manner.

{The decoder $E_d$ also has a ResNet50-like backbone with four layers, except a VIN module appended after each residual block in the decoder. Each layer of the decoder consists of two components, the residual block and the VIN, which produce the corresponding output feature maps $r_k$ and $h_k$, $k\in \left\{1,2,3,4\right\}$. In the residual block, the spatial resolution of the feature map $h_{k+1}$ is first up-sampled by a factor of 2. Then the up-sampled feature is concatenated with the feature $f_k$ from the encoder and compressed by a $1\times1$ convolution to reduce channel dimensions. The compressed feature map is further processed by a $3\times3$ convolution and generates the output feature map of the residual block $r_k$:}
\begin{equation}
	r_k = conv3\left(conv1\left(concate\left(up\_sample(h_{k+1}),f_k\right)\right)\right),
	\end{equation}
where $up\_sample$, $conv1$, $conv3$ denote up-sampling, $1\times1$, and $3\times3$ operations, respectively. Note that $r_4=conv3\left(f_4\right)$.

\noindent\textbf{Variational information normalization (VIN)}.  
As shown in Fig.~\ref{fig:Framework} (c), the VIN module takes the feature map $r_k$ and the guidance of vector $\mathbf{v}_c$ as inputs, and outputs an HQ-aware feature map $h_k$ that is combined with the decoded feature map $r_k$ for further processing.
VIN is a residual connection structure with AdaLIN, LReLU, and a $3\times3$ convolution layer. Therefore, the final feature map  $h_k$ is formulated as follows:
    \begin{equation}
    \begin{aligned}
        h_k =& V_{IN}\left(r_k, \mathbf{v}_c\right),\\
        =& conv3\left(AdaLIN\left(r_k, \mathbf{v}_c\right)\right),
    \end{aligned}
	\end{equation}
where $AdaLIN$ is the AdaLIN operator, which is an information normalization block by combining the adaptive instance normalization (AdaIN)~\cite{huang2017arbitrary} and the adaptive layer normalization (AdaLN)~\cite{zhang2022styleswin} with a learnable parameter. The definition of the proposed AdaLIN is presented as follows:
	\begin{equation}
		AdaLIN = \rho AdaIN + (1-\rho) AdaLN,
	\end{equation}
	where $\rho$ is a learnable parameter and is initialized as 0.5. Although AdaIN and AdaLN focus on instance and layer separately, they share a common structure with the same mathematical formulation. Take AdaIN as an example:
\begin{equation}	AdaIN\left(r_k,\textbf{v}_c\right)=\sigma\left(\textbf{v}_c\right)\left(\frac{r_k-\mu(r_k)}{\sigma\left(r_k\right)}\right)+\mu\left(\textbf{v}_c\right),
\label{eq:AdaIN}	
\end{equation}
    where $\mu$ and $\sigma$ correspond to mean and variance value. {Being normalized by the concise and general variational information from the joint vector $\textbf{v}_c$ (including $\textbf{v}_z$ and $\textbf{v}_y$), HQG-Net can simultaneously exploit the HQ cues extracted from the vector-based HQ guidance and preserve significant structural information from the LQ input. This ensures the generation of enhancement results with visual fidelity and structure preservation, thereby facilitating clinical decision-making.}
	
\noindent\textbf{Discriminator}. {Our discriminator $E_{dis}$} is constructed following PatchGAN~\cite{li2016precomputed} with patch-wise strategy, which is particularly beneficial for the LQ medical images with complex degraded conditions~\cite{deng2022pcgan}, e.g., blur and retinal artifact~\cite{zhao2020automated}, and thus ensures the enhancement performance.

\subsection{Content-Aware Loss} 
Previous enhancement losses focus on the pixel-level consistency on both paired settings, e.g., mean square error, and unpaired conditions, e.g., cycle-consistency loss~\cite{ma2021structure}, for the structure preservation and visual fidelity, which brings two problems for UMIE. First, it can be challenging for pixel-level constraints to distinguish the complete structural information between the concerning biological tissues and LQ factors, especially for complex degradations. Therefore, a preliminary structure extraction operation can benefit the pixel-level loss function. Besides, the pixel-level constraint suffers from limited information, whereas in-depth features have a more robust representation capacity. Hence, an additional feature-level constraint is desirable to supplement this deficiency.

% and thus can compensate for this deficiency. Consequently, it is desirable to propose an extra feature-level constraint to compensate for the representational deficit.
	
Since structural information is predominantly encoded in high-frequency components~\cite{ma2016infrared} while exhibiting the texture sparsity in the whole image space, it is preferable to encourage structure fidelity in separately extracted high-frequency components rather than the whole-frequency image space. The wavelet-based operator is well-suited to this requirement \cite{yoo2019photo,he2023camouflaged}. To this end, we adopt the Haar wavelet to construct a structure preservation loss, termed $\mathcal{L}_{Haar}$, between the LQ input $\mathbf{y}$ and the enhanced result $\widehat{\mathbf{x}}$. $\mathcal{L}_{Haar}$ is defined as follows:
\begin{equation}\label{eq:HaarLoss}
\mathcal{L}_{Haar}=\left\|\widehat{\mathbf{x}}^{HF}-\mathbf{y}^{HF}\right\|_1,
\end{equation}
	{where $\widehat{\mathbf{x}}^{HF}$ and $\mathbf{y}^{HF}$ are the integrated high-frequency parts of $\widehat{\mathbf{x}}$ and $\mathbf{y}$ filtered by the wavelet-based high pass filters, including the vertical, horizontal, and diagonal filters. $\mathcal{L}_{Haar}$ enforce the network to learn how to utilize the extracted HQ cues without sacrificing structure preservation from the LQ input, thus reducing the risk of misleading medical decisions.}
	
For the second problem, a feature consistency loss, dubbed $\mathcal{L}_{FC}$, is proposed by constraining the feature-level consistency between the enhanced result $\widehat{\mathbf{x}}$ and the comprehensive guidance $\textbf{v}_c$ in a vector form, which is formulated as:
\begin{equation} \label{Eq:FCLoss}
\mathcal{L}_{FC}=\left\|Gr(\mathbf{v}_{\widehat{\mathbf{x}}})-Gr(\textbf{v}_c)\right\|_F,
	\end{equation}
	where $\mathbf{v}_{\widehat{\mathbf{x}}}=concate\left(G(\widehat{\mathbf{x}}),FC_e\left(E_{e}(\mathbf{\widehat{\mathbf{x}}})\right)\right)$, $Gr(\bigcdot)$ represents the Gram matrix~\cite{kacem2018novel} for a more abstract feature characterization. $\|\bigcdot\|_F$ denotes Frobenius norm. As presented in Eq.~\ref{Eq:FCLoss}, the enhanced result $\widehat{\mathbf{x}}$ is expected to share a similar representation with the generated encoding information of the LQ input and the variational encoding feature of the HQ guidance in a compact vector form. {Such a constraint contributes to the extraction of general HQ cues and ensures that the extracted HQ cues can be fully leveraged for image enhancement, thereby jointly ensuring the retention of feature-level information and visual fidelity.}
	
	Combining the Haar wavelet-based loss and feature consistency loss, we propose a content-aware loss, $\mathcal{L}_{CA}$, for structure preservation, textural enhancement, and visual fidelity:
	\begin{equation}
		\begin{aligned}
		\mathcal{L}_{CA}&=\mathcal{L}_{Haar}+\lambda_1 \mathcal{L}_{FC}, \\
		&=\left\|\widehat{\mathbf{x}}^{HF}-\mathbf{y}^{HF}\right\|_1+\lambda_1\left\|Gr(\mathbf{v}_{\widehat{\mathbf{x}}})-Gr(\textbf{v}_c)\right\|_F,
		\end{aligned}
	\end{equation}
	where $\lambda_1$ is a trade-off parameter. Intuitively, the content-aware loss balances the structural preservation from LQ input $\mathbf{y}$, and the feature-based visual fidelity from HQ guidance $\mathbf{z}$. 
	
	Apart from the content-aware loss, a standard GAN loss is used for the adversarial training to simultaneously regularize the generation ability and discrimination capability:
	\begin{equation}
		\begin{aligned}
		\mathcal{L}_{GAN}&=\underset{E_{gen}}\min \ \underset{E_{dis}}\max \ \mathcal{L}(E_{gen},E_{dis}), \\
			&=\mathbb{E}_{\mathbf{y}}\left[\log E_{dis}(\mathbf{y})\right]+\mathbb{E}_{\mathbf{y},\mathbf{z}}\left[\log (1- E_{dis}(\widehat{\mathbf{x}})) \right],
		\end{aligned}
	\end{equation}
	where $E_{gen}(\bigcdot)$ generates an enhanced result $\widehat{\mathbf{x}}$ conditioned on the LQ image $\mathbf{y}$ and the guidance image $\mathbf{z}$, while ${E_{dis}}(\bigcdot)$ tries to distinguish the differences between LQ and HQ images.
	
Combining the content-aware loss and the standard GAN loss, we reach our enhancement-oriented learning objective as
	\begin{equation}
	    \mathcal{L}^{e}=\mathcal{L}_{CA}+\lambda_{2}\mathcal{L}_{GAN},
	\end{equation}
where $\lambda_{2}$ is the trade-off parameter.
	
\subsection{Bi-level Optimization and Cooperative Training}
\begin{figure}
		\centering
		\includegraphics[width=0.5\textwidth]{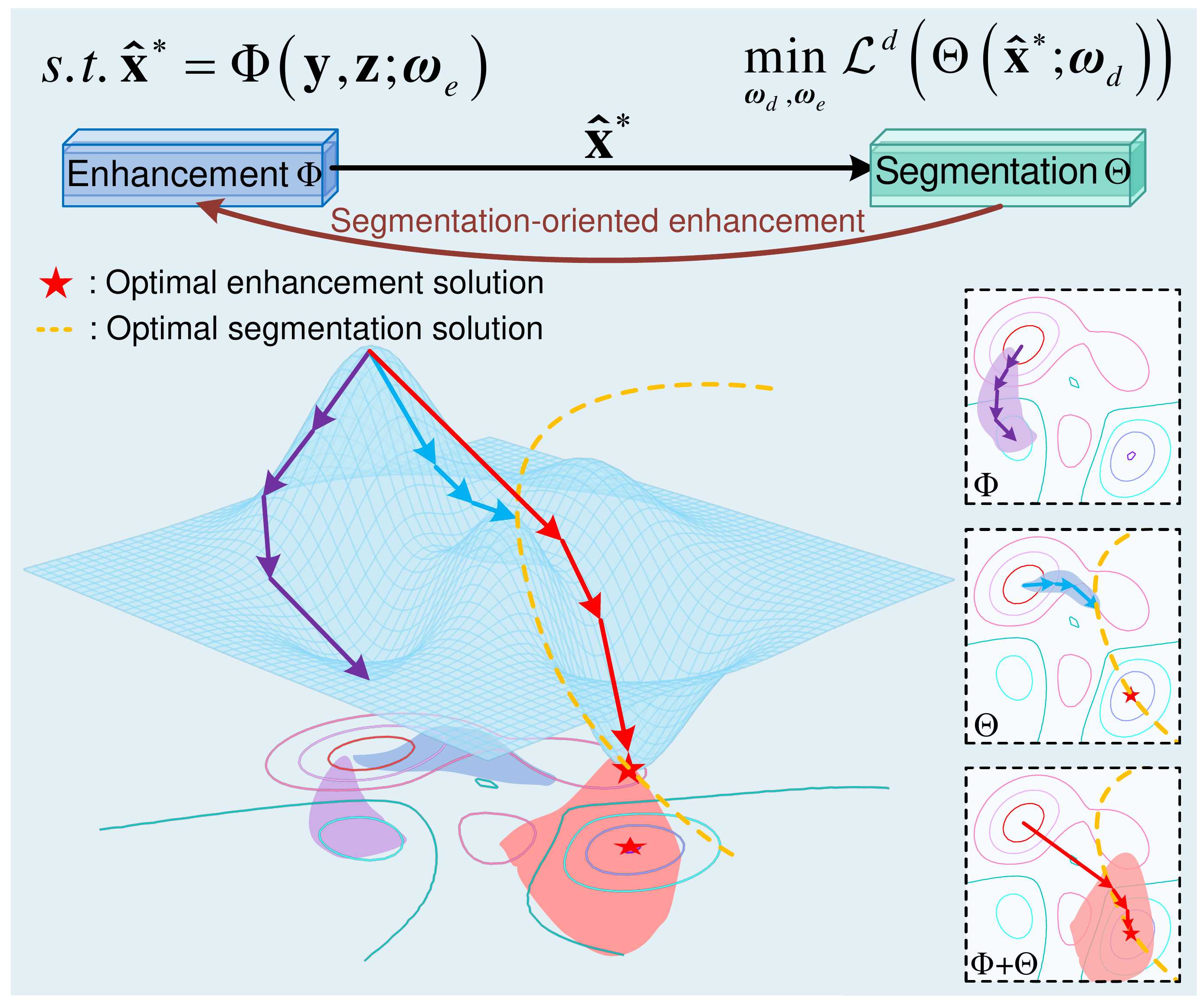}
				\vspace{-5mm}
		\caption{Bi-level optimization formulation and cooperative training strategy, where the right three contour plots denote the optimization processes constrained by the enhancement  $\Phi$, the segmentation $\Theta$, and the bi-level strategy $\Phi+\Theta$.} 
		\label{fig:Bi-level}
		\vspace{-4mm}
	\end{figure}
	
\noindent\textbf{Bi-level optimization}. Due to discrepancies in domain knowledge and training strategy, there inevitably exists a significant domain gap between low-level vision problems and subsequent high-level tasks, i.e., a high-quality reconstruction result in the low-level domain may not necessarily result in a good performance in high-level tasks~\cite{ju2022ivf}. Inspired by the success of bi-level optimization (BLO) in parameter optimization that jointly updates the model parameters and the hyper-parameters~\cite{liu2022task}, we introduce BLO to UMIE, aiming to achieve satisfactory performance on both the UMIE problem and downstream tasks. Following the truism Stackelberg's theory~\cite{ochs2015bilevel}, the segmentation-oriented enhancement is defined as a BLO formulation, which is formulated as follows:
		\begin{equation}
				\underset{\bm{\omega}_d}{\min} \ \mathcal{L}^{d}\left(\Theta\left(\widehat{\mathbf{x}}^{*};\bm{\omega}_d\right)\right), 
		\end{equation}
		\begin{equation} \label{eq:Bi-levelConstraint}
			s.t. \ \widehat{\mathbf{x}}^{*}\in \arg \underset{\widehat{\mathbf{x}}}{\min} f\left(\widehat{\mathbf{x}};\mathbf{y},\mathbf{z}\right)+g_e\left(\widehat{\mathbf{x}};\mathbf{y}\right)+g_v\left(\widehat{\mathbf{x}};\mathbf{z}\right),
		\end{equation}
	where $\mathcal{L}^{d}$ is the objective function for the segmentation task and $\Theta\left(\bigcdot;\bm{\omega}_d\right)$ represents the segmentation network enabled by the trainable parameters $\bm{\omega}_d$. $f\left(\bigcdot\right)$ is the enhancement term jointly driven by the two feasible constraints $g_e\left(\bigcdot\right)$ and $g_v\left(\bigcdot\right)$.
	
As presented in Fig.~\ref{fig:Bi-level}, the proposed bi-level formulation can improve the performance of both enhancement and segmentation tasks. Paradoxically, the ill-posed constraint in Eq.~\ref{eq:Bi-levelConstraint} is not bound in the form of an equation, which inevitably brings challenges for further optimization. Consequently, it is desirable to replace the complicated constraints with an adaptive framework and unroll the bi-level formulation into two networks, i.e., the segmentation network $\Theta$ and the enhancement network $\Phi$, which is defined as follows:
	\begin{equation}\label{Eq:Bi2Single}
		\underset{\bm{\omega}_d, \bm{\omega}_e}{\min} \ \mathcal{L}^{d}\left(\Theta\left(\widehat{\mathbf{x}}^{*};\bm{\omega}_d\right)\right), \ s.t. \ \widehat{\mathbf{x}}^{*}=\Phi\left(\mathbf{y},\mathbf{z};\bm{\omega}_e\right),
	\end{equation}
where $\Phi\left(\bigcdot;\bm{\omega}_e\right)$ is the proposed HQG-Net with the corresponding parameters $\bm{\omega}_e$. In practice, considering the discrepancy of medical data and medical applications, different deep frameworks are adopted as the backbone for the downstream network $\Theta$ under the specific task. In particular, CS-Net~\cite{mou2019cs} and PNS-Net~\cite{ji2021progressively} are correspondingly applied as the segmentation network for \textit{CCM} and \textit{colonoscopy} datasets, and Deepgrading~\cite{mou2022deepgrading} are used as the classification network for nerve fiber tortuosity classification task with CCM data.
	
\noindent\textbf{Cooperative training strategy}. In this section, we further propose a cooperative training strategy for the jointly optimal network parameters, i.e., $\bm{\omega}=\left(\bm{\omega}_d,\bm{\omega}_e\right)$, with the aforementioned bi-level framework. By incorporating the downstream regularizer $\mathcal{L}^d$ and the enhancement regularizer $\mathcal{L}^e$, Eq.~\ref{Eq:Bi2Single} can be rewritten from the perspective of mutual optimization:
\begin{equation} \label{Eq:CoopTrain}
	\underset{\bm{\omega}_d, \bm{\omega}_e}{\min} \ \mathcal{L}^{d}\left(\Theta\left(\widehat{\mathbf{x}}^{*};\bm{\omega}_d\right)\right)+\lambda_{3}\mathcal{L}^{e}\left(\Phi\left(\mathbf{y},\mathbf{z};\bm{\omega}_e\right)\right), \ 
\end{equation}
\begin{equation} \label{Eq:CoopCons}
	s.t. \ \widehat{\mathbf{x}}^{*}=\Phi\left(\mathbf{y},\mathbf{z};\bm{\omega}_e\right),
\end{equation}
where $\lambda_{3}$ is a hyperparameter for balance. We further present the gradient propagation flow of the parameters of downstream task $\bm{\omega}_d$ and enhancement network $\bm{\omega}_e$:
\begin{equation}
	\frac{\partial\mathcal{L}}{\partial\bm{\omega}_d}=\frac{\partial\mathcal{L}^{d}}{\partial\Theta_d}\frac{\partial\Theta_d}{\partial\bm{\omega}_d}, 
\end{equation}
\begin{equation}
        \frac{\partial\mathcal{L}}{\partial\bm{\omega}_e}=\frac{\partial\mathcal{L}^{d}}{\partial\Theta_d}\frac{\partial\Theta_d}{\partial\Phi_e}\frac{\partial\Phi_e}{\partial\bm{\omega}_e}+\lambda_{3}\frac{\partial\mathcal{L}^{e}}{\partial\Phi_e}\frac{\partial\Phi_e}{\partial\bm{\omega}_e},
\end{equation}
where $\mathcal{L}$ denotes the overall energy function Eq.~\ref{Eq:CoopTrain}, which is subject to Eq.~\ref{Eq:CoopCons}. 
{The loss gradient of $\bm{\omega}_e$ is jointly determined by the loss functions of both enhancement and the downstream task, yielding visually pleasant results with favorable downstream performance.}
\begin{figure*}[htbp!]
				\centering
				\includegraphics[width=1\textwidth]{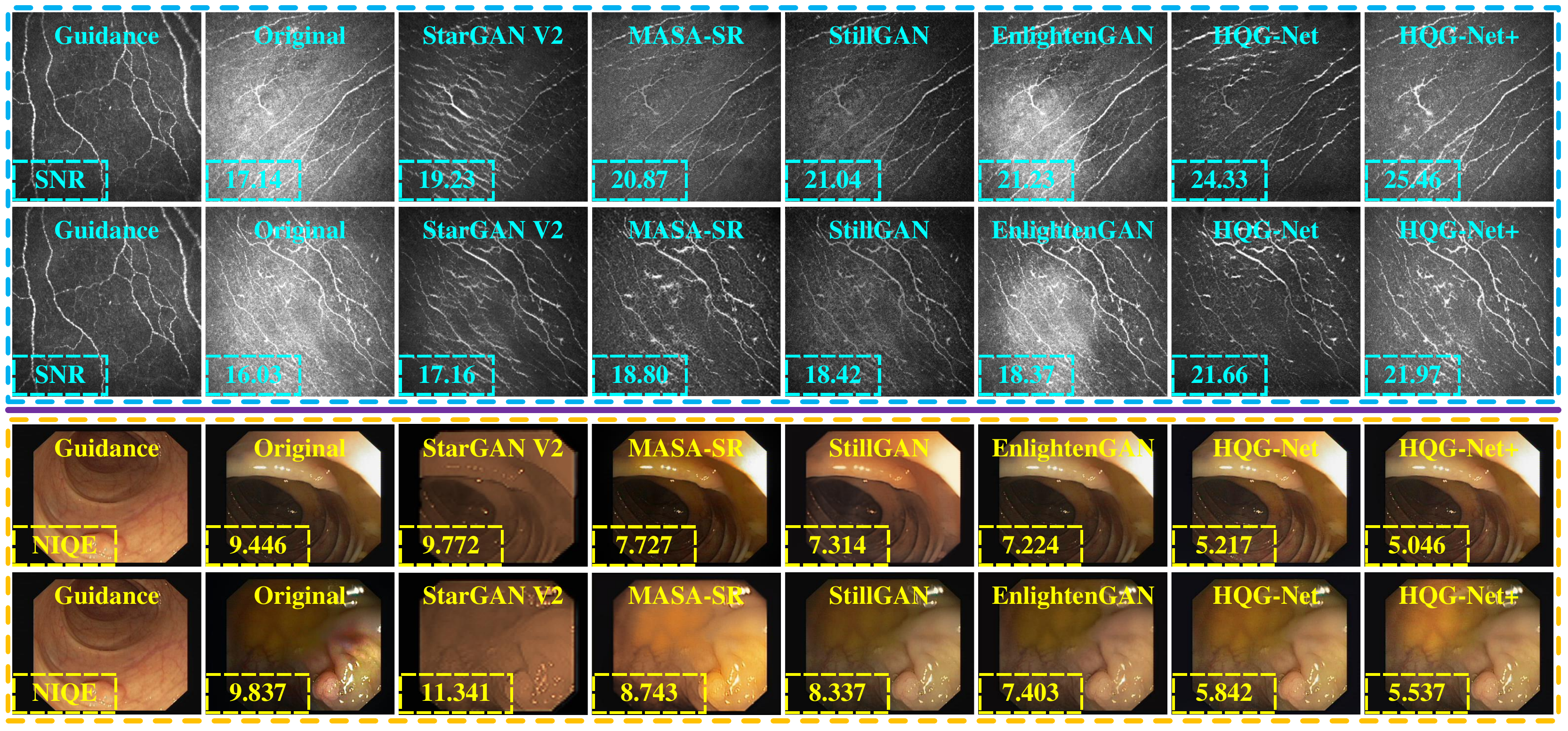}
				\caption{Visual comparison of enhanced results on \textit{CCM} and \textit{Colonoscopy} datasets with guidance images. $SNR$ is $SNR_{r=3}$.}
				\label{fig:CCM_CVC_Sup}
				\vspace{-0.3cm}
			\end{figure*}
\begin{table}[htbp!]
		\centering
		\caption{Details of the used datasets. \textit{CCM} and \textit{Colonoscopy} datasets have the paired segmented labels.}\vspace{-2mm}
		\label{table:dataset}
		%\scalebox{1}{
			%\resizebox{\columnwidth}{!}{
				\setlength{\tabcolsep}{2.5mm}
				\begin{tabular}{l|cc|cc|c}
					\toprule[1.5pt]
					\multirow{2}{*}{Datasets} & \multicolumn{2}{c|}{Training Set} & \multicolumn{2}{c|}{Test Set}            & \multirow{2}{*}{Paired Labels}  \\ \cline{2-5} 
					& \multicolumn{1}{c|}{HQ}    & LQ  & \multicolumn{1}{c|}{HQ} & LQ                   \\ \hline \hline
					\textit{CCM}  & \multicolumn{1}{c|}{288}   & 340 &\multicolumn{1}{c|}{200} & 60                 & \checkmark            \\ \hline
					\textit{Colonoscopy}               & \multicolumn{1}{c|}{115}   & 270 &\multicolumn{1}{c|}{200} & 70       & \checkmark             \\ \hline
					\textit{Fundus }                    & \multicolumn{1}{c|}{640}   & 700 &\multicolumn{1}{c|}{200} & 600                &               \\ \bottomrule[1.5pt]
		\end{tabular}
		\vspace{-4mm}
	\end{table}
{Note that under the cooperative training strategy, the enhancement task is not unilaterally catered to the downstream task. Conversely, the downstream task can also assist the enhancement task to obtain better enhanced results.
For instance, in medical image segmentation, mis-segmentation often occurs at the boundaries that contain crucial structural information and texture details, which are also challenging to highlight for UMIE. In this regard,
% Take the medical image segmentation task as an example. The segmentation task separates the image into the foreground and the background, and the mis-segmentation often exists at the boundaries, which also contain structural information and texture details that are difficult to highlight for the UMIE task. In this case, 
the enhancement network, with additional constraints on the loss function of the segmentation task, can focus more on the boundary information and thus generate visually pleasant results with preserved structure and enhanced texture. Hence, as shown in Fig.~\ref{fig:Bi-level}, compared to the enhancement network optimized solely with the enhancement loss, the network optimized by our cooperative training strategy with a BLO formulation can achieve a more optimal enhancement solution with the assistance of constraints from the downstream task.}
	
\section{Experiment}
\noindent\textbf{Datasets.} {We employ three datasets, i.e., the corneal confocal microscopy (\textit{CCM}) dataset, the \textit{Fundus} dataset, and the \textit{Colonoscopy} dataset, to evaluate enhancement performance under complex degeneration conditions. $\textit{CCM}$ dataset is publicly available \cite{ma2021structure}, while \textit{Fundus} and \textit{Colonoscopy} datasets are the private datasets collected and relabelled by our collaborative clinicians into HQ and LQ subsets from the iSee dataset~\cite{yan2019oversampling} and the CVC-EndoSceneStill dataset~\cite{vazquez2017benchmark}. Details of the three datasets are presented in Table \ref{table:dataset}. Note that \textit{CCM} and \textit{Colonoscopy}  contain paired segmentation labels for the HQ and LQ images, which enables us to quantitatively evaluate the enhancement quality by taking segmentation as the downstream task and retrain our framework with the proposed cooperative training strategy for a bi-level optimization.} 
	
		\begin{table*}[tbp!]
		\centering
		\caption{Quantitative evaluation of enhancement quality. 
				The best and second-best results are marked in red and blue, respectively. ``+'' denotes optimizing the method under the cooperative training strategy. In \textit{CCM} dataset, the cooperative training strategy improves the enhancement result by $6.9\%$ (HQG-Net), $2.5\%$ (EnlightenGAN), $4.9\%$ (StillGAN), $5.3\%$ (MASA-SR), and $1.2\%$ (StarGAN V2). In \textit{Colonoscopy} dataset, the cooperative training strategy increases that by $8.7\%$ (HQG-Net), $5.6\%$ (EnlightenGAN), $3.7\%$ (StillGAN), $2.9\%$ (MASA-SR), and $15.0\%$ (StarGAN V2).}
			\label{table:enhancement}
			\scalebox{1}{
				% \resizebox{2\columnwidth}{!}{
					%\setlength{\tabcolsep}{2mm}
					%\setlength{\tabcolsep}{2mm}{
						\begin{tabular}{l|cccc|cccc}
							\toprule[1.5pt]
							\multicolumn{1}{l|}{}  	& \multicolumn{4}{c|}{\textit{CCM}}                                        & \multicolumn{4}{c}{\textit{Colonoscopy}}                                \\ \cline{2-9}
							\multicolumn{1}{l|}{\multirow{-2}{*}{Methods}}      & \multicolumn{1}{c|}{\cellcolor{gray!40}$SNR_{r=3}$ $\uparrow$}&\multicolumn{1}{c|}{\cellcolor{gray!40}$SNR_{r=5}$ $\uparrow$}&\multicolumn{1}{c|}{\cellcolor{gray!40}$SNR_{r=7}$ $\uparrow$}&\cellcolor{gray!40}$SNR_{r=9}$ $\uparrow$& \multicolumn{1}{c|}{\cellcolor{gray!40}AG $\uparrow$}& \multicolumn{1}{c|}{\cellcolor{gray!40}EN $\uparrow$}&\multicolumn{1}{c|}{\cellcolor{gray!40}NIQE $\downarrow$}                        & \cellcolor{gray!40}BRISQUE $\downarrow$                      \\ \hline \hline
							Original     & \multicolumn{1}{c|}{17.47}&\multicolumn{1}{c|}{17.61}&\multicolumn{1}{c|}{17.65}&17.67&\multicolumn{1}{c|}{1.10}&\multicolumn{1}{c|}{4.29}&\multicolumn{1}{c|}{12.89}& 51.76                        \\ \hline
							CAEFI~\cite{jeelani2019content} &\multicolumn{1}{c|}{18.12}&\multicolumn{1}{c|}{18.97}&\multicolumn{1}{c|}{19.73}&21.35&\multicolumn{1}{c|}{4.33}&\multicolumn{1}{c|}{6.32}&\multicolumn{1}{c|}{6.82}& 39.46                        \\ \hline
							ATA~\cite{zhao2020automated} &\multicolumn{1}{c|}{14.27}&\multicolumn{1}{c|}{15.07}&\multicolumn{1}{c|}{17.72}& 18.83&\multicolumn{1}{c|}{5.84}&\multicolumn{1}{c|}{6.18}                        & \multicolumn{1}{c|}{6.43}                        & 44.73\\ \hline
							IDRLP~\cite{ju2021idrlp} & \multicolumn{1}{c|}{15.14}                        &\multicolumn{1}{c|}{17.54}&\multicolumn{1}{c|}{20.57}& 21.86&\multicolumn{1}{c|}{2.59}&\multicolumn{1}{c|}{5.91}                       & \multicolumn{1}{c|}{6.85}                        & 42.07\\ \hline
							StarGAN V2~\cite{choi2020stargan} & \multicolumn{1}{c|}{19.44} &\multicolumn{1}{c|}{19.62}&\multicolumn{1}{c|}{19.76}&20.73&\multicolumn{1}{c|}{3.16}&\multicolumn{1}{c|}{5.97}                        & \multicolumn{1}{c|}{7.41}                        & 41.83\\ \hline
                            SSEN~\cite{shim2020robust} & \multicolumn{1}{c|}{19.12} &\multicolumn{1}{c|}{19.52}&\multicolumn{1}{c|}{19.93}&21.04&\multicolumn{1}{c|}{5.96}&\multicolumn{1}{c|}{5.89}                        & \multicolumn{1}{c|}{9.43}                        & 40.45\\ \hline
                            TTSR~\cite{yang2020learning} & \multicolumn{1}{c|}{20.78} &\multicolumn{1}{c|}{21.32}&\multicolumn{1}{c|}{22.04}&22.93&\multicolumn{1}{c|}{6.33}&\multicolumn{1}{c|}{6.05}                        & \multicolumn{1}{c|}{7.57}                        & 39.70\\ \hline
                            $C^2$-Matching~\cite{jiang2021robust} & \multicolumn{1}{c|}{20.34} &\multicolumn{1}{c|}{20.85}&\multicolumn{1}{c|}{21.73}&21.96&\multicolumn{1}{c|}{6.07}&\multicolumn{1}{c|}{5.78}                        & \multicolumn{1}{c|}{7.94}                        & 39.86\\ \hline
                            MASA-SR~\cite{lu2021masa} & \multicolumn{1}{c|}{\color[HTML]{00B0F0} \textbf{21.23}} &\multicolumn{1}{c|}{21.78}&\multicolumn{1}{c|}{22.31}&22.94&\multicolumn{1}{c|}{6.50}&\multicolumn{1}{c|}{6.17}                        & \multicolumn{1}{c|}{7.02}                        & \color[HTML]{00B0F0} \textbf{39.05}\\ \hline
							StillGAN~\cite{ma2021structure} & \multicolumn{1}{c|}{{20.45}} & \multicolumn{1}{c|}{21.11}& \multicolumn{1}{c|}{21.94}&22.74                        &\multicolumn{1}{c|}{6.32}& \multicolumn{1}{c|}{6.56}&\multicolumn{1}{c|}{\color[HTML]{00B0F0}\textbf{6.62}}&\multicolumn{1}{c}{{39.42}}                         \\ \hline
							EnlightenGAN~\cite{jiang2021enlightengan} & \multicolumn{1}{c|}{20.22}                        &\multicolumn{1}{c|}{{\color[HTML]{00B0F0} \textbf{22.87}}}&\multicolumn{1}{c|}{{\color[HTML]{00B0F0} \textbf{23.55}}}&{\color[HTML]{00B0F0} \textbf{24.37}}&\multicolumn{1}{c|}{{\color[HTML]{00B0F0} \textbf{6.60}}}&\multicolumn{1}{c|}{{\color[HTML]{00B0F0} \textbf{6.85}}}& \multicolumn{1}{c|}{{6.69}}                        & 41.15\\ \hline
							HQG-Net         & \multicolumn{1}{c|}{{\color[HTML]{FF0000} \textbf{22.68}}}               & \multicolumn{1}{c|}{{\color[HTML]{FF0000} \textbf{23.19}}}               & \multicolumn{1}{c|}{{\color[HTML]{FF0000} \textbf{25.08}}}               & {{\color[HTML]{FF0000} \textbf{26.14}}}               & \multicolumn{1}{c|}{{{\color[HTML]{FF0000} \textbf{6.75}}}}               & \multicolumn{1}{c|}{{{\color[HTML]{FF0000} \textbf{7.26}}}}               & \multicolumn{1}{c|}{{\color[HTML]{FF0000} \textbf{6.61}} }              & {{\color[HTML]{FF0000}\textbf{ 36.43}} }                \\ \hline \hline
                            %再加上StarGAN和一个能提升效果的SR（最好是MASA-SR）
                            StarGAN V2+ & \multicolumn{1}{c|}{19.55} &\multicolumn{1}{c|}{19.78}&\multicolumn{1}{c|}{20.35}&20.84&\multicolumn{1}{c|}{5.26}&\multicolumn{1}{c|}{6.33}                        & \multicolumn{1}{c|}{7.02}                        & 38.83\\ \hline
                            MASA-SR+ & \multicolumn{1}{c|}{22.37} &\multicolumn{1}{c|}{23.06}&\multicolumn{1}{c|}{23.41}&24.10&\multicolumn{1}{c|}{6.54}&\multicolumn{1}{c|}{6.33}                        & \multicolumn{1}{c|}{7.01}                        & 38.94\\ \hline
							StillGAN+ & \multicolumn{1}{c|}{21.74} & \multicolumn{1}{c|}{22.35} & \multicolumn{1}{c|}{22.87} & \multicolumn{1}{c|}{23.42}&\multicolumn{1}{c|}{6.38}&\multicolumn{1}{c|}{6.67}&\multicolumn{1}{c|}{6.54}&\multicolumn{1}{c}{39.33}\\ \hline
							EnlightenGAN+ & \multicolumn{1}{c|}{21.03} & \multicolumn{1}{c|}{23.28} & \multicolumn{1}{c|}{23.97} & \multicolumn{1}{c|}{24.93}&\multicolumn{1}{c|}{6.68}&\multicolumn{1}{c|}{6.87}&\multicolumn{1}{c|}{6.68}&\multicolumn{1}{c}{39.31}\\ \hline
                            HQG-Net+ & \multicolumn{1}{c|}{24.13}                        &\multicolumn{1}{c|}{26.24}&\multicolumn{1}{c|}{26.87}& 27.12&\multicolumn{1}{c|}{6.93}&\multicolumn{1}{c|}{6.92}                       & \multicolumn{1}{c|}{6.43}                        &33.17\\ \bottomrule[1.5pt]
				\end{tabular}}
    % }
				\vspace{-2mm}
			\end{table*}

	\noindent\textbf{Implementation details.}	Our HQG-Net is trained with random flipping for data augmentation.  Adam optimizer is applied with momentum terms $(0.5,0.99)$ and the learning rate is set as $1\times10^{-4}$. The batch size is set as 4, and the trade-off parameters $\lambda_{1}$, $\lambda_{2}$, $\lambda_{3}$ are set as 10, 1, 5, respectively. In the test phase, one HQ image in the corresponding test set is randomly selected as guidance for the LQ input. All the experiments are implemented with PyTorch on two RTX3090TI GPUs.

\begin{figure*}
		\centering
		\includegraphics[width=0.9\textwidth]{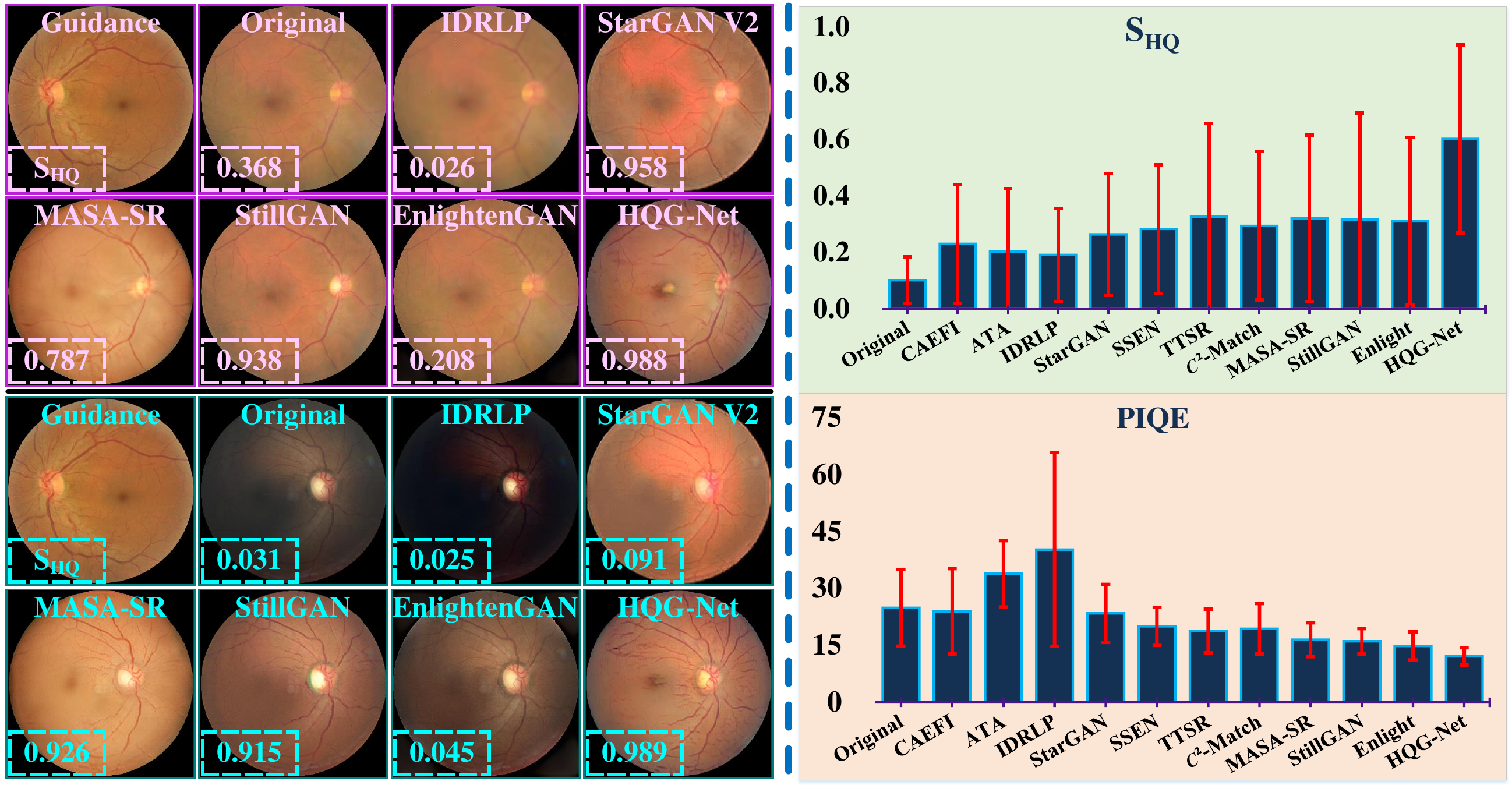}
		\vspace{-3mm}
		\caption{Quanlitative and quantitative comparison of the enhanced results on fundus dataset, where StarGAN, $C^2$-Match, Enlight are short for StarGAN V2, $C^2$-Matching, EnlightenGAN. A larger value in $S_{HQ}$ means better performance, while that is reversed in PIQE. The $S_{HQ}$ of the HQ guidance is 0.986.} \vspace{-2mm}
		\label{fig:Fundus-Enhancement}
		\vspace{-2mm}
	\end{figure*}

 \begin{table*}[t]
		\centering
		\caption{Parameters (M) and FLOPs (G) with the image size of $384\times384$ (both the LQ input and the HQ guidance).%, where HQG-Net achieves the lowest FLOPs.
        %, where the best and the second-best are marked in red and blue.
        }\vspace{-1mm}
		\label{table:runningtime}
		\scalebox{1}{
			\resizebox{2\columnwidth}{!}{
            \setlength{\tabcolsep}{2.9mm}
				\begin{tabular}{l|c|c|c|c|c|c|c|c}
					\toprule[1.5pt]
					& \cellcolor{c2!50}StarGAN V2\cite{choi2020stargan}&\cellcolor{c2!50}SSEN~\cite{shim2020robust}& \cellcolor{c2!50}TTSR~\cite{yang2020learning}&\cellcolor{c2!50}$C^2$-Matching~\cite{jiang2021robust}& \cellcolor{c2!50}MASA-SR~\cite{lu2021masa} & \cellcolor{c2!50}StillGAN\cite{ma2021structure}& \cellcolor{c2!50}EnlightenGAN\cite{jiang2021enlightengan}& \cellcolor{c2!50}HQG-Net  \\ \hline \hline
					%CCM (S)       & 8.731 & 3.271 & 2.016 & 4.354      & {\color[HTML]{00B0F0} \textbf{1.344}}    & 2.032        & {\color[HTML]{FF0000}\textbf{1.312}} \\ \hline
					%Fundus (S)     & 7.223 & 2.365 & 1.253 & 2.377      & {\color[HTML]{00B0F0} \textbf{1.171}}    & 1.287        & {\color[HTML]{FF0000}\textbf{1.101}} \\ \hline
					%Colonscopy (S) & 9.302 & 3.551 & 2.201 & 5.075      & {\color[HTML]{00B0F0} \textbf{1.481}}    & 2.342        & {\color[HTML]{FF0000}\textbf{1.404}} \\ \hline
					Params (M)&78.95&6.52&7.04&9.68&4.38 &{{58.38}} &{{56.37}} & 66.42\\ \hline
					FLOPs (G)&259.32&285.34&556.41&300.25&258.04 &{{94.73}} &98.67 &{{87.33}} \\ \bottomrule[1.5pt]
		\end{tabular}}}
		\vspace{-1mm}
	\end{table*}

 \begin{table*}[t]
    \caption{Ablation studies. $SNR$ is $SNR_{r=3}$. \textit{Colon} is short for \textit{Colonoscopy}. ``w/'' and ``w/o'' denote with and without. ``A$\rightarrow$B'' means substitute A for B. In (b), const is short for constraints. $L_1$ and $L_2$ denote $L_{InterVar}$ and $L_{IntraVar}$. (e) is conducted on the combined feature maps
    %from $f_4$ and the HQ cue extraction module 
    rather than the joint vector $\textbf{v}_c$ for feature matching. (*,*,*) in (g) means whether using the HF components in vertical, horizontal, and diagonal directions, e.g., (1,0,0) indicates using HF components in the vertical direction while discarding those in horizontal and diagonal directions. %The best results are marked in \textbf{bold}. 
    }\vspace{-1mm}
    \label{table:AblationStudy}
    \begin{subtable}{.352\textwidth}
		\centering
        \caption{Effect of the HQ guidance mechanism.}\vspace{-2mm}
		\label{table:EffectHQ}
    \resizebox{1\columnwidth}{!}{
    \setlength{\tabcolsep}{0.4mm}
    \begin{tabular}{l|l|c|c|c}\toprule[1.5pt]
Data& Metrics  &\cellcolor{c2!50} w/o guidance & \cellcolor{c2!50}Learnable tensor & \cellcolor{c2!50}w/ guidance \\ \hline\hline
\multirow{2}{*}{\textit{CCM}} & $SNR$ $\uparrow$   & 18.62           & 19.83            & \textbf{22.68}          \\ \cline{2-5}
& Dice $\uparrow$    & 0.47            & 0.48  & \textbf{0.54}   \\ \hline
\multirow{2}{*}{\textit{Fundus}}      & $S_{HQ}$ $\uparrow$& 0.28            & 0.34             & \textbf{0.60}          \\ \cline{2-5}
& PIQE $\downarrow$    & 19.92           & 17.64            & \textbf{12.16} \\ \hline
\multirow{2}{*}{\textit{Colon}} & NIQE $\downarrow$    & 7.32& 7.05             & \textbf{6.61}           \\ \cline{2-5}
& Dice $\uparrow$    & 0.87  & 0.89   & \textbf{0.91} \\ \bottomrule[1.5pt] 
\end{tabular}}\vspace{1mm}
    \end{subtable}
    \begin{subtable}{.335\textwidth}
		\centering
        \caption{Consistency of the guidance representation.}\vspace{-2mm}
		\label{table:CompactGuidance}
    \resizebox{1\columnwidth}{!}{
    \setlength{\tabcolsep}{0.65mm}
    \begin{tabular}{l|l|c|c|c|c}\toprule[1.5pt]
Data& Metrics  &\cellcolor{c2!50} w/o const &\cellcolor{c2!50} w/ $L_1$ &\cellcolor{c2!50} w/ $L_2$ &\cellcolor{c2!50} w/ $L_1$+$L_2$ \\ \hline\hline
\multirow{2}{*}{\textit{CCM}}& $SNR$ $\uparrow$& 22.68 & 22.67 & \textbf{22.69} & 22.67 \\ \cline{2-6}
& Dice $\uparrow$ & \textbf{0.54} &\textbf{ 0.54} & \textbf{0.54} & \textbf{0.54} \\ \hline
\multirow{2}{*}{\textit{Fundus}}& $S_{HQ}$ $\uparrow$& \textbf{0.60}            & \textbf{0.60} & \textbf{0.60} & \textbf{0.60} \\ \cline{2-6}
& PIQE $\downarrow$ & \textbf{12.16} & 12.18 & 12.25 & 12.23 \\ \hline
\multirow{2}{*}{\textit{Colon}} & NIQE $\downarrow$ & 6.61 & 6.64   & \textbf{6.60} & 6.65 \\ \cline{2-6}
& Dice $\uparrow$& \textbf{0.91}& 0.90& 0.90& 0.90\\ \bottomrule[1.5pt]    
\end{tabular}}\vspace{1mm}\end{subtable} 
    \begin{subtable}{.30\textwidth}
		\centering
        \caption{Robustness to the HQ guidance.}\vspace{-2mm}
		\label{table:RobustHQ}
    \resizebox{1\columnwidth}{!}{
    \setlength{\tabcolsep}{0.4mm}
    \begin{tabular}{l|l|c|c|c} \toprule[1.5pt]
Data                     & Metrics   & \cellcolor{c2!50}All data & \cellcolor{c2!50}10 images &\cellcolor{c2!50} 1 fixed image \\ \hline\hline
\multirow{2}{*}{\textit{CCM}}         & $SNR$ $\uparrow$    & 22.56           & \textbf{22.71}          & 22.68\\ \cline{2-5}
& Dice $\uparrow$     & \textbf{0.54}            & 0.53     & \textbf{0.54}           \\ \hline
\multirow{2}{*}{\textit{Fundus}}      & $S_{HQ}$ $\uparrow$ & \textbf{0.61}            & 0.60      & 0.60\\ \cline{2-5}
& PIQE $\downarrow$     & \textbf{11.93}           & 12.70     & 12.16          \\ \hline
\multirow{2}{*}{\textit{Colon}} & NIQE $\downarrow$     & 6.63            & 6.52      & \textbf{6.61} \\ \cline{2-5}
& Dice $\uparrow$     & 0.90 & \textbf{0.91}  & \textbf{0.91} \\ \bottomrule[1.5pt]    
\end{tabular}}\vspace{1mm}
    \end{subtable}\\ 
    \begin{subtable}{.498\textwidth}
		\flushleft
        \caption{Effect of the joint vector $\mathbf{v}_z$ and AdaLIN.}\vspace{-2mm}
		\label{table:EffectAdaLIN}
    \resizebox{1\columnwidth}{!}{
    \setlength{\tabcolsep}{0.4mm}
    \begin{tabular}{l|l|c|c|c|c|c} \toprule[1.5pt]
Data & Metrics  & \cellcolor{c2!50}w/o multiscale & \cellcolor{c2!50}$\mathbf{v}_z$$\rightarrow$$\mathbf{v}_c$ & \cellcolor{c2!50}AdaIN$\rightarrow$AdaLIN & \cellcolor{c2!50}AdaLN$\rightarrow$AdaLIN& \cellcolor{c2!50}w/ AdaLIN \\ \hline \hline
& $SNR$ $\uparrow$   & 21.15 & 20.73& 22.12& 21.72& \textbf{22.68} \\ \cline{2-7}
\multirow{-2}{*}{\textit{CCM}}& Dice $\uparrow$&0.51 & 0.51& 0.52 & 0.50 & \textbf{0.54} \\\hline
& $S_{HQ}$ $\uparrow$& 0.52& 0.53& 0.56& 0.53& \textbf{0.60} \\ \cline{2-7}
\multirow{-2}{*}{\textit{Fundus}}& PIQE $\downarrow$& 16.73& 17.97& 14.14& 13.72& \textbf{12.16}    \\\hline
 & NIQE $\downarrow$    & 7.74           & 7.82                    & 7.14                       & 6.90 & \textbf{6.61} \\\cline{2-7}
\multirow{-2}{*}{\textit{Colon}} & Dice $\uparrow$    & 0.84           & 0.85                    & 0.90                       & 0.88                       & \textbf{0.91}  \\ \bottomrule[1.5pt]  
\end{tabular}}\vspace{1mm}
    \end{subtable} \hspace{1mm}
    \begin{subtable}{.487\textwidth}
		\flushright
        \caption{AdaLIN \textit{vs.} other integration strategies.}\vspace{-2mm}
		\label{table:AdaLINVersus}
    \resizebox{1\columnwidth}{!}{
    \setlength{\tabcolsep}{0.4mm}
    \begin{tabular}{l|l|c|c|c|c} \toprule[1.5pt]
Data  & Metrics  & \cellcolor{c2!50}Concatenate$\rightarrow$AdaLIN & \cellcolor{c2!50}Add$\rightarrow$AdaLIN & \cellcolor{c2!50}Multiple$\rightarrow$AdaLIN & \cellcolor{c2!50}w/ AdaLIN\\ \hline\hline
& $SNR$ $\uparrow$   & 20.04  & 20.10 & 19.86& \textbf{22.19} \\ \cline{2-6}
\multirow{-2}{*}{\textit{CCM}}    & Dice $\uparrow$    & 0.51& 0.51 & 0.50  & \textbf{0.52}                          \\ \hline
& $S_{HQ}$ $\uparrow$& 0.50   & 0.49 & 0.46   & \textbf{0.55} \\\cline{2-6}
\multirow{-2}{*}{\textit{Fundus}} & PIQE $\downarrow$    & 16.38 & 17.06 & 16.82 & \textbf{12.94}  \\ \hline
& NIQE $\downarrow$    & 7.28  & 7.55& 7.73 & \textbf{6.92} \\\cline{2-6}
\multirow{-2}{*}{\textit{Colon}} & Dice $\uparrow$    & 0.85 & 0.84 & 0.84  & \textbf{0.87}   \\ \bottomrule[1.5pt]  
\end{tabular}}\vspace{1mm}
    \end{subtable} \\ 
    \begin{subtable}{.465\textwidth}
		\flushleft
        \caption{Ablation study of the content-aware loss $L_{CA}$.}\vspace{-2mm}
		\label{table:AblationCAloss}
    \resizebox{1\columnwidth}{!}{
    \setlength{\tabcolsep}{0.6mm}
    \begin{tabular}{l|l|c|c|c|c|c}\toprule[1.5pt]
Data                     & Metrics  & \cellcolor{c2!50}w/o $L_{CA}$ &\cellcolor{c2!50} $L_{Haar}$$\rightarrow$$L_{CA}$ & \cellcolor{c2!50}$L_{FC}$$\rightarrow$$L_{CA}$ & \cellcolor{c2!50}$L_s$$\rightarrow$$L_{CA}$ &\cellcolor{c2!50} w/ $L_{CA}$ \\ \hline\hline
& $SNR$ $\uparrow$   & 21.16                          & 21.45                 & 21.29               & 21.82                & \textbf{22.68}        \\ \cline{2-7}
\multirow{-2}{*}{\textit{CCM}}    & Dice $\uparrow$    & 0.48                           & 0.53                  & 0.49                & 0.51                 & \textbf{0.54}         \\ \hline
& $S_{HQ}$ $\uparrow$& 0.40              & 0.48                  & 0.54                & 0.53                 & \textbf{0.60}    \\ \cline{2-7}
\multirow{-2}{*}{\textit{Fundus}} & PIQE $\downarrow$    & 16.30                          & 15.40                 & 13.76               & 14.81                & \textbf{12.16}        \\ \hline
& NIQE $\downarrow$    & 8.71     & 7.25    & 7.13                & 7.36                 & \textbf{6.61}         \\ \cline{2-7}
\multirow{-2}{*}{\textit{Colon}} & Dice  $\uparrow$   & 0.80                           & 0.87                  & 0.89                & 0.83                 & \textbf{0.91}    \\ \bottomrule[1.5pt]  
\end{tabular}}\vspace{-4mm}
    \end{subtable}\hspace{2mm}
    \begin{subtable}{.52\textwidth}
		\centering
        \caption{Different combinations of high-frequency components in $L_{Haar}$.}\vspace{-2mm}
		\label{table:CombinationHF}
    \resizebox{1\columnwidth}{!}{
    \setlength{\tabcolsep}{2mm}
    \begin{tabular}{l|l|c|c|c|c|c|c|c}\toprule[1.5pt]
Data& Metrics  & \cellcolor{c2!50}(1,0,0) & \cellcolor{c2!50}(0,1,0) & \cellcolor{c2!50}(0,0,1) & \cellcolor{c2!50}(1,1,0) & \cellcolor{c2!50}(1,0,1) & \cellcolor{c2!50}(0,1,1) & \cellcolor{c2!50}(1,1,1) \\ \hline\hline
& $SNR$ $\uparrow$   & 21.54& 21.49   & 22.03   & 22.14   & 22.44   & 22.46   & \textbf{22.68}   \\ \cline{2-9}
\multirow{-2}{*}{CCM}    & Dice $\uparrow$    & 0.52                           & 0.51    & 0.52    & 0.52    & 0.53    & 0.53    & \textbf{0.54}    \\ \hline
& $S_{HQ}$ $\uparrow$ & 0.46 & 0.47    & 0.52    & 0.53    & 0.57    & 0.56    & \textbf{0.60}    \\ \cline{2-9}
\multirow{-2}{*}{Fundus} & PIQE $\downarrow$    & 15.37  & 15.42   & 14.07   & 13.85   & 12.94   & 12.86   & \textbf{12.16}   \\ \hline
 & NIQE $\downarrow$    & 7.86 & 7.92    & 7.73    & 7.46    & 7.12    & 6.97    & \textbf{6.61}    \\ \cline{2-9}
\multirow{-2}{*}{Colono} & Dice $\uparrow$    & 0.83 & 0.82    & 0.83    & 0.86    & 0.89    & 0.90    & \textbf{0.91}   \\ \bottomrule[1.5pt] 
\end{tabular}}\vspace{-4mm}
    \end{subtable}
\end{table*}

\noindent\textbf{Compared methods.} {Ten state-of-the-art (SOTA) methods are selected in comparison, including three traditional methods, i.e., CARFI~\cite{jeelani2019content}  (Retinex-based), ATA~\cite{zhao2020automated} (Retinex-based), and IDRLP~\cite{ju2021idrlp} (region line prior), and seven learning-based techniques, i.e., StarGAN V2~\cite{choi2020stargan} (style transfer), SSEN~\cite{shim2020robust} (ref-SR), TTSR~\cite{yang2020learning} (ref-SR), $C^2$-Matching~\cite{jiang2021robust} (ref-SR), MASA-SR~\cite{lu2021masa} (ref-SR), StillGAN~\cite{ma2021structure} (CycleGAN-based), and EnlightenGAN~\cite{jiang2021enlightengan} (Pix2Pix-based). Note that we incorporate the style transfer-oriented or ref-SR-based techniques in natural scenarios into our experiment to evaluate the effectiveness of the task-specific modifications of our HQG-Net mentioned in Section~\ref{sec:relatedwork}. All the compared methods are trained and tested with their corresponding default settings. }

\subsection{Results and Analyses}
\noindent\textbf{CCM.} {All the learning-based methods are trained on the $\textit{CCM}$ dataset~\cite{ma2021structure}. Four metrics, i.e., signal-to-noise ratio (SNR)~\cite{ma2021structure} with a radius of 3, 5, 7, and 9 pixels, are used for quantitative evaluation, where a higher score indicates a better enhancement result. As shown in Table~\ref{table:enhancement}, the proposed HQG-Net achieves the best performance in all the metrics with a significant improvement. Furthermore, we also provide the results of StarGAN V2+, MASA-SR+, StillGAN+, EnlightenGAN+, and HQG-Net+, where + means the network is trained with the cooperative training strategy, and the higher metric scores obtained by all the re-trained networks indicate the generalization and effectiveness of the cooperative strategy.

The visualization results are presented in Fig.~\ref{fig:CCM_CVC_Sup}, where we only present partially enhanced results for space limitation. As shown in Fig.~\ref{fig:CCM_CVC_Sup}, CAEFI fails to enhance the LQ image with good visual fidelity. StarGAN V2 utilizes the same guidance image as our method, while the generated images suffer from texture blur and local luminance distortion, which lies in that the style transfer technique has neither an information maintenance framework nor a degradation-specific loss function. MASA-SR also fails to preserve critical structures and can be influenced by complex degradation due to its ignorance of the characteristics of medical images.} StillGAN fails to enhance texture information and highlight the contrast for lacking the direct involvement of real HQ images in the enhancement, as well as the coarse structure loss. Global illumination uniformity is a problem for EnlightenGAN for the excessive attention to local details. In contrast, the enhanced results from HQG-Net enjoy high contrast and enhanced texture details, which is mainly attributed to the explicit HQ information guided HQG-Net and the content-aware loss function. {In addition, the results of HQG-Net+ better conform to the definitions of HQ images in Fig.~\ref{fig:HQDefinition}. Specifically, the LQ CCM data enhanced by HQG-Net+ can better enhance the texture details than those enhanced by HQG-Net, which owes to the segmentation-oriented cooperative training strategy.}
\begin{table*}[t]
    \centering
    \caption{LQ data enhanced by HQG-Net under the guidance of different data, including medical data and natural data. Medical data are from the three used datasets, i.e., \textit{Fundus}, \textit{CCM}, and \textit{Colonoscopy}. Natural data include the hazy data from RESIDE~\cite{li2018benchmarking}, low-light data from SID~\cite{chen2018learning}, noisy data from SIDD~\cite{abdelhamed2018high}, and the HQ data from their ground-truth images. 
    } \label{table:RobustAnaGuidance}
    \begin{subtable}{.53\textwidth}
		\centering
        \caption{CCM data guided by different medical data.}\vspace{-2mm}
		\label{table:RobustAnaCCMMedical}
    \resizebox{1\columnwidth}{!}{
    \setlength{\tabcolsep}{0.6mm}
    \begin{tabular}{l|cccccc}
    \toprule[1.5pt]
     Metrics& \multicolumn{1}{c|}{\cellcolor{c2!50}HQ \textit{CCM}} & \multicolumn{1}{c|}{\cellcolor{c2!50}HQ \textit{Fundus}} & \multicolumn{1}{c|}{\cellcolor{c2!50}HQ \textit{Colonoscopy}} & \multicolumn{1}{c|}{\cellcolor{c2!50}LQ \textit{CCM}} & \multicolumn{1}{c|}{\cellcolor{c2!50}LQ \textit{Fundus}} & \multicolumn{1}{c}{\cellcolor{c2!50}LQ \textit{Colonoscopy}} \\ \hline \hline
$SNR$ $\uparrow$ & \multicolumn{1}{c|}{\textbf{22.68}} & \multicolumn{1}{c|}{21.14} & \multicolumn{1}{c|}{21.08} & \multicolumn{1}{c|}{19.76} & \multicolumn{1}{c|}{19.15} & \multicolumn{1}{c}{19.33} \\ \hline
Dice $\uparrow$     & \multicolumn{1}{c|}{\textbf{0.54}} & \multicolumn{1}{c|}{0.52} & \multicolumn{1}{c|}{0.51} & \multicolumn{1}{c|}{0.49} & \multicolumn{1}{c|}{0.47} & \multicolumn{1}{c}{0.48} \\ \bottomrule[1.5pt]
\end{tabular}}
    \end{subtable}
    \begin{subtable}{.462\textwidth}
		\centering
        \caption{CCM data guided by different natural data.}\vspace{-2mm}
		\label{table:RobustAnaCCMNatural}
    \resizebox{1\columnwidth}{!}{
    \setlength{\tabcolsep}{0.6mm}
    \begin{tabular}{l|cccc}
    \toprule[1.5pt]
         Metrics& \multicolumn{1}{c|}{\cellcolor{c2!50}HQ natural data} & \multicolumn{1}{c|}{\cellcolor{c2!50}Hazy data~\cite{li2018benchmarking}} & \multicolumn{1}{c|}{\cellcolor{c2!50}Low-light data~\cite{chen2018learning}} & \multicolumn{1}{c}{\cellcolor{c2!50}Noisy data~\cite{abdelhamed2018high}} \\ \hline \hline
$SNR$ $\uparrow$ & \multicolumn{1}{c|}{19.07} & \multicolumn{1}{c|}{18.82} & \multicolumn{1}{c|}{18.73} & \multicolumn{1}{c}{\textbf{19.11}} \\ \hline
Dice $\uparrow$  & \multicolumn{1}{c|}{\textbf{0.47}} & \multicolumn{1}{c|}{0.46} & \multicolumn{1}{c|}{\textbf{0.47}} & \multicolumn{1}{c}{\textbf{0.47}}  \\ \bottomrule[1.5pt]    \end{tabular}}
    \end{subtable}\vspace{2mm}\\
    \begin{subtable}{.53\textwidth}
		\centering
        \caption{Fundus data guided by different medical data.}\vspace{-2mm}
		\label{table:RobustAnaFundusMedical}
    \resizebox{1\columnwidth}{!}{
    \setlength{\tabcolsep}{0.6mm}
    \begin{tabular}{l|cccccc}
    \toprule[1.5pt]
     %multirow{2}{*}{\textit{Fundus}} & \multicolumn{6}{c}{{Under the guidance with different medical data}}\\ \cline{2-7}
     Metrics& \multicolumn{1}{c|}{\cellcolor{c2!50}HQ \textit{CCM}} &\multicolumn{1}{c|}{\cellcolor{c2!50}HQ \textit{Fundus}} & \multicolumn{1}{c|}{\cellcolor{c2!50}HQ \textit{Colonoscopy}} & \multicolumn{1}{c|}{\cellcolor{c2!50}LQ \textit{CCM}}& \multicolumn{1}{c|}{\cellcolor{c2!50}LQ \textit{Fundus}}  & \multicolumn{1}{c}{\cellcolor{c2!50}LQ \textit{Colonoscopy}} \\ \hline \hline
$S_{HQ}$ $\uparrow$ & \multicolumn{1}{c|}{0.47} & \multicolumn{1}{c|}{\textbf{0.60}} & \multicolumn{1}{c|}{0.49} & \multicolumn{1}{c|}{0.29}& \multicolumn{1}{c|}{0.35}  & \multicolumn{1}{c}{0.29} \\ \hline
PIQE $\downarrow$      & \multicolumn{1}{c|}{14.75}& \multicolumn{1}{c|}{\textbf{12.16}} & \multicolumn{1}{c|}{13.82}& \multicolumn{1}{c|}{19.84} & \multicolumn{1}{c|}{17.36}  & \multicolumn{1}{c}{18.72} \\ \bottomrule[1.5pt]
\end{tabular}}
    \end{subtable}
    \begin{subtable}{.462\textwidth}
		\centering
        \caption{Fundus data guided by different natural data.}\vspace{-2mm}
		\label{table:RobustAnaFundusNatural}
    \resizebox{1\columnwidth}{!}{
    \setlength{\tabcolsep}{0.6mm}
    \begin{tabular}{l|cccc}
    \toprule[1.5pt]
         %\multirow{2}{*}{\textit{Fundus}}& \multicolumn{4}{c}{{ Under the guidance with different natural data}}\\ \cline{2-5}
         Metrics& \multicolumn{1}{c|}{\cellcolor{c2!50}HQ natural data} & \multicolumn{1}{c|}{\cellcolor{c2!50}Hazy data~\cite{li2018benchmarking}} & \multicolumn{1}{c|}{\cellcolor{c2!50}Low-light data~\cite{chen2018learning}} & \multicolumn{1}{c}{\cellcolor{c2!50}Noisy data~\cite{abdelhamed2018high}} \\ \hline \hline
$S_{HQ}$ $\uparrow$ & \multicolumn{1}{c|}{\textbf{0.29}} & \multicolumn{1}{c|}{\textbf{0.29}} & \multicolumn{1}{c|}{0.28} & \multicolumn{1}{c}{\textbf{0.29}} \\ \hline
PIQE $\downarrow$  & \multicolumn{1}{c|}{19.65} & \multicolumn{1}{c|}{\textbf{19.43}} & \multicolumn{1}{c|}{20.07} & \multicolumn{1}{c}{19.89}  \\ \bottomrule[1.5pt]    \end{tabular}}
    \end{subtable}\vspace{2mm}\\
    \begin{subtable}{.53\textwidth}
		\centering
        \caption{Colonoscopy data guided by different medical data.}\vspace{-2mm}
		\label{table:RobustAnaColoMedical}
    \resizebox{1\columnwidth}{!}{
    \setlength{\tabcolsep}{0.6mm}
    \begin{tabular}{l|cccccc}
    \toprule[1.5pt]
     Metrics& \multicolumn{1}{c|}{\cellcolor{c2!50}HQ \textit{CCM}} & \multicolumn{1}{c|}{\cellcolor{c2!50}HQ \textit{Fundus}}& \multicolumn{1}{c|}{\cellcolor{c2!50}HQ \textit{Colonoscopy}}   & \multicolumn{1}{c|}{\cellcolor{c2!50}LQ \textit{CCM}} & \multicolumn{1}{c|}{\cellcolor{c2!50}LQ \textit{Fundus}}& \multicolumn{1}{c}{\cellcolor{c2!50}LQ \textit{Colonoscopy}} \\ \hline \hline
$NIQE$ $\downarrow$  & \multicolumn{1}{c|}{6.89} & \multicolumn{1}{c|}{6.83} & \multicolumn{1}{c|}{\textbf{6.61}} & \multicolumn{1}{c|}{7.31} & \multicolumn{1}{c|}{7.12}& \multicolumn{1}{c}{6.96} \\ \hline
Dice $\uparrow$      & \multicolumn{1}{c|}{0.90} & \multicolumn{1}{c|}{0.90}& \multicolumn{1}{c|}{\textbf{0.91}}  & \multicolumn{1}{c|}{0.87} & \multicolumn{1}{c|}{0.88}& \multicolumn{1}{c}{0.89} \\ \bottomrule[1.5pt]
\end{tabular}}\vspace{-4mm}
\end{subtable}
\begin{subtable}{.462\textwidth}
		\centering
        \caption{Colonoscopy data guided by different natural data.}\vspace{-2mm}
		\label{table:RobustAnaColoNatural}
    \resizebox{1\columnwidth}{!}{
    \setlength{\tabcolsep}{0.6mm}
    \begin{tabular}{l|cccc}
    \toprule[1.5pt]
         %\multirow{2}{*}{\textit{Fundus}}& \multicolumn{4}{c}{{ Under the guidance with different natural data}}\\ \cline{2-5}
         Metrics& \multicolumn{1}{c|}{\cellcolor{c2!50}HQ natural data} & \multicolumn{1}{c|}{\cellcolor{c2!50}Hazy data~\cite{li2018benchmarking}} & \multicolumn{1}{c|}{\cellcolor{c2!50}Low-light data~\cite{chen2018learning}} & \multicolumn{1}{c}{\cellcolor{c2!50}Noisy data~\cite{abdelhamed2018high}} \\ \hline \hline
$NIQE$ $\downarrow$ & \multicolumn{1}{c|}{7.41} & \multicolumn{1}{c|}{\textbf{7.17}} & \multicolumn{1}{c|}{7.40} & \multicolumn{1}{c}{7.46} \\ \hline
Dice $\uparrow$  & \multicolumn{1}{c|}{0.87} & \multicolumn{1}{c|}{\textbf{0.88}} & \multicolumn{1}{c|}{0.86} & \multicolumn{1}{c}{0.87}  \\ \bottomrule[1.5pt]    \end{tabular}} \vspace{-4mm}
    \end{subtable}
\end{table*}

\noindent\textbf{Colonoscopy.} We train all the learning-based methods with the private \textit{Colonoscopy} dataset and evaluate the enhancement performance with four metrics, i.e., average gradient (AG)~\cite{wu2005remote}, entropy (EN)~\cite{haussler1997mutual}, natural image quality evaluator (NIQE)~\cite{mittal2012making}, and blind/reference image spatial quality evaluator (BRISQUE)~\cite{mittal2012no}. A higher value in AG or EN means better results, which is reversed in NIQE and BRISQUE.  Table~\ref{table:enhancement} exhibits the qualitative results, and the proposed HQG-Net has the best performance on all the metrics. {Besides, we also demonstrate the superiority of the cooperative training strategy in terms of its capacity to improve enhancement performance.} Fig.~\ref{fig:CCM_CVC_Sup} also demonstrates the superiority of HQG-Net in visual comparison, where our enhanced results have excellent visual fidelity for our variation-guided framework and the task-specific loss function. {Moreover, in Fig.~\ref{fig:CCM_CVC_Sup}, profiting from the cooperative training strategy, the enhanced results in HQG-Net+ have more uniform illumination than those in HQG-Net.}

\noindent\textbf{Fundus.} \textit{Fundus} dataset is used for training the compared networks with 640 HQ images and 700 LQ images. High-quality score ($S_{HQ}$)~\cite{ma2021structure} and perception-based image quality evaluator (PIQE)~\cite{zhang2015feature} are used for evaluation, where a higher value in $S_{HQ}$ or a lower score in PIQE means a better outcome. The qualitative and quantitative analyses are presented in Fig.~\ref{fig:Fundus-Enhancement}. In qualitative analysis, we achieve the best performance in $S_{HQ}$ and PIQE, where our $S_{HQ}$ is 0.64 and is almost twice as high as the second-best result.  Furthermore, the qualitative analysis also demonstrates that the enhanced fundus image has uniform illumination, enhanced texture details, and excellent visual fidelity. Note that the cooperative training strategy is not applicable to this task for the lack of paired labels.
	
\noindent\textbf{Computational efficiency.} {We compare the parameters of deep networks and the corresponding FLOPs (at the size of $384\times384$) on the enhancement task in Table~\ref{table:runningtime}. As shown in Table~\ref{table:runningtime}, the proposed HQG-Net achieves the lowest FLOPs, which demonstrates the efficiency of our HQG-Net.}

\begin{figure*}[htbh!]
	\centering
	\begin{subfigure}{0.163\textwidth}
		\centering
		\includegraphics[width=\textwidth]{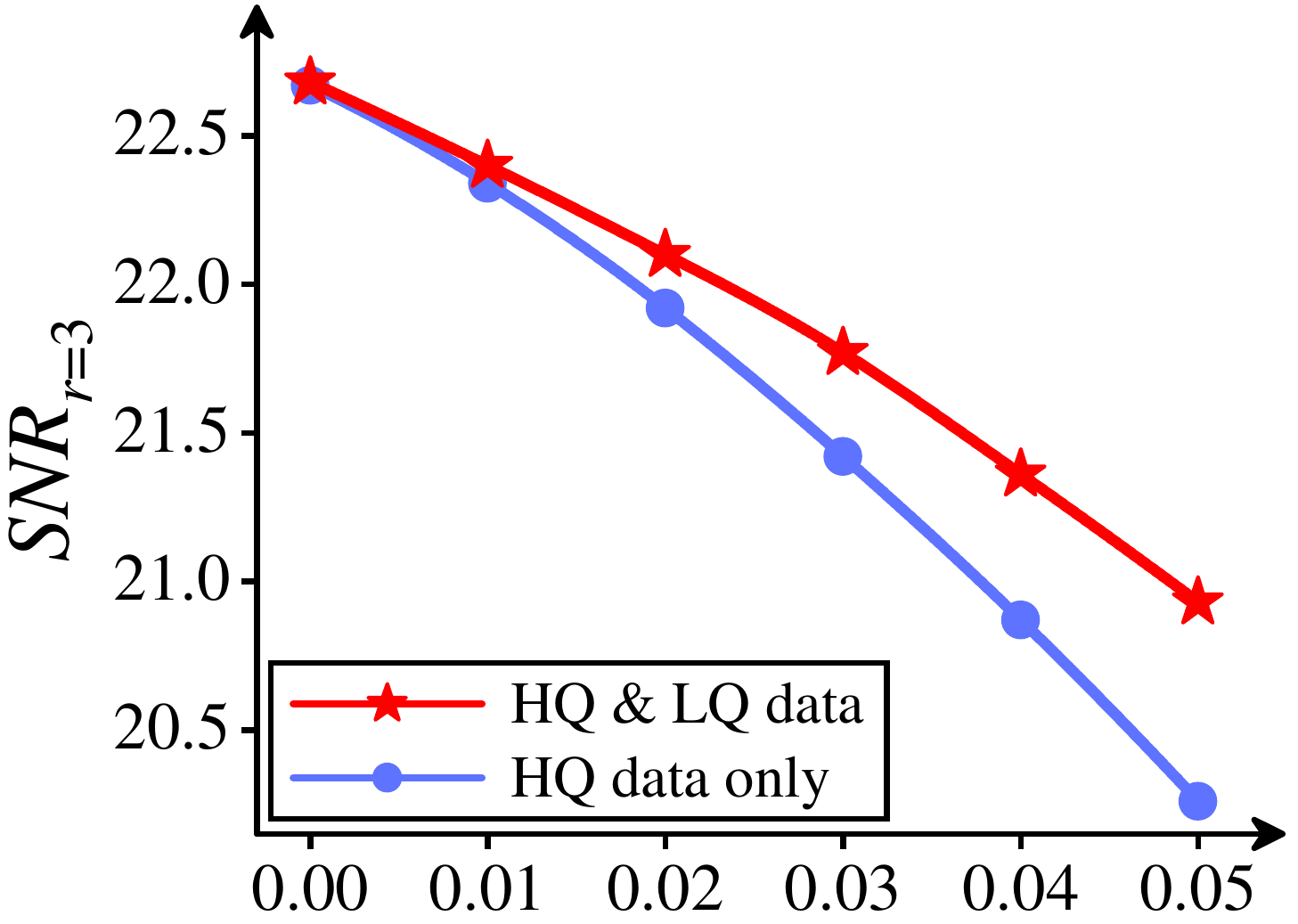}\vspace{-2pt}
		\caption{\footnotesize Gaussian noise.}
	\end{subfigure}
	\hfill
	\begin{subfigure}{0.163\textwidth}  
		\centering 
		\includegraphics[width=\textwidth]{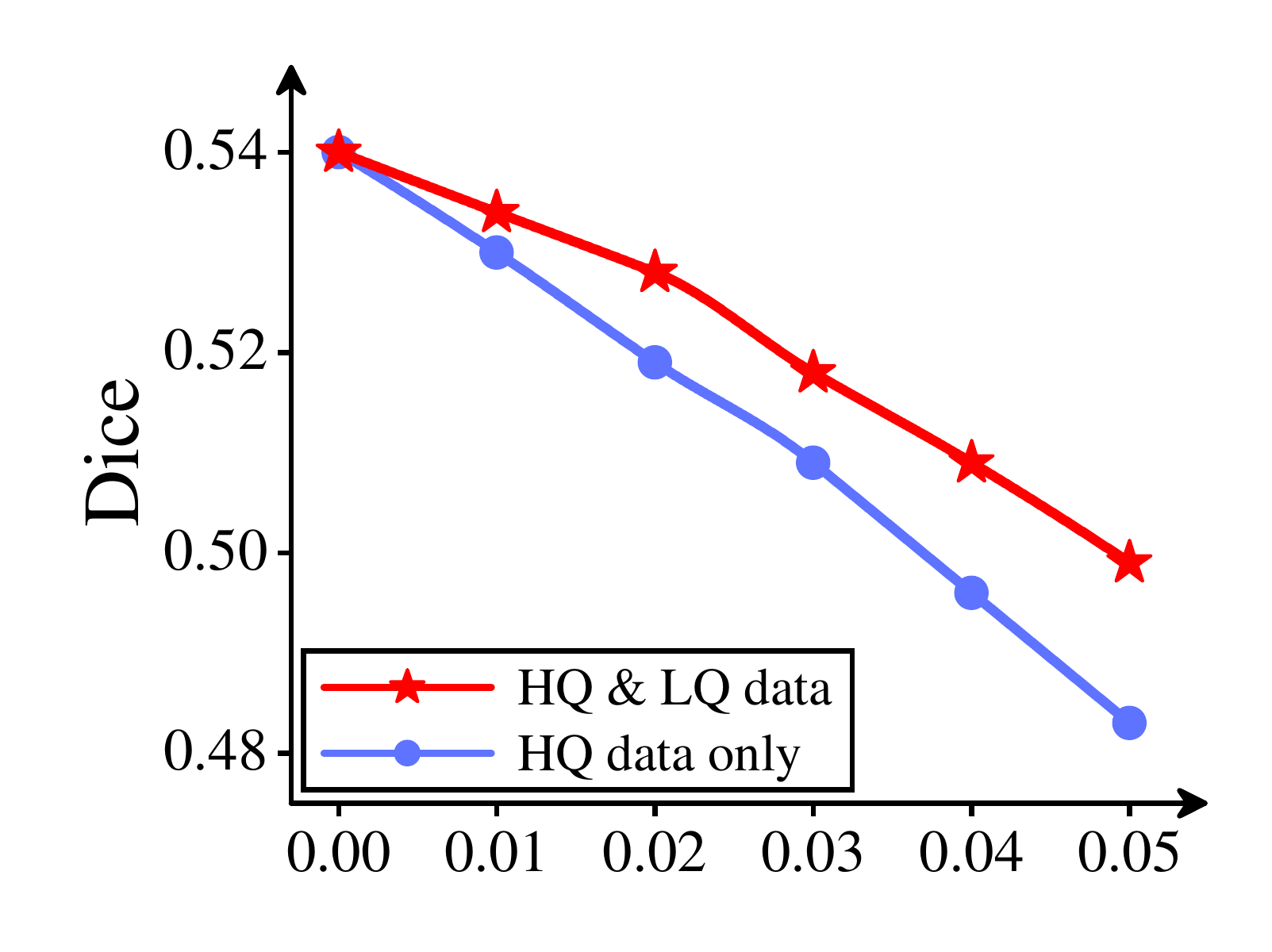}\vspace{-2pt}
		\caption{\footnotesize Gaussian noise.}
	\end{subfigure}%\vspace{-1mm}
    \begin{subfigure}{0.163\textwidth}
		\centering
		\includegraphics[width=\textwidth]{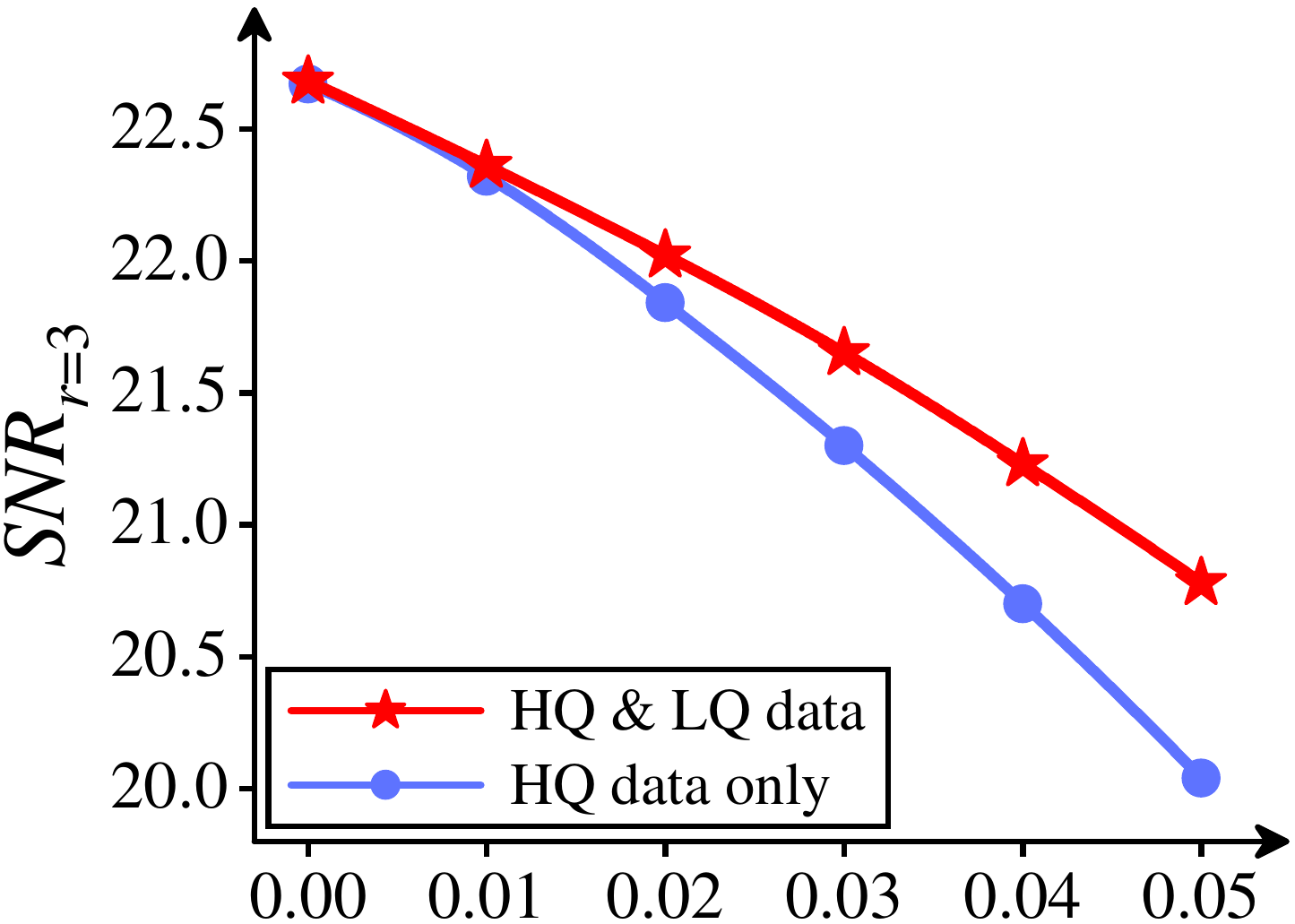}\vspace{-2pt}
		\caption{\footnotesize Poisson noise.}
	\end{subfigure}
	\hfill
	\begin{subfigure}{0.163\textwidth}  
		\centering 
		\includegraphics[width=\textwidth]{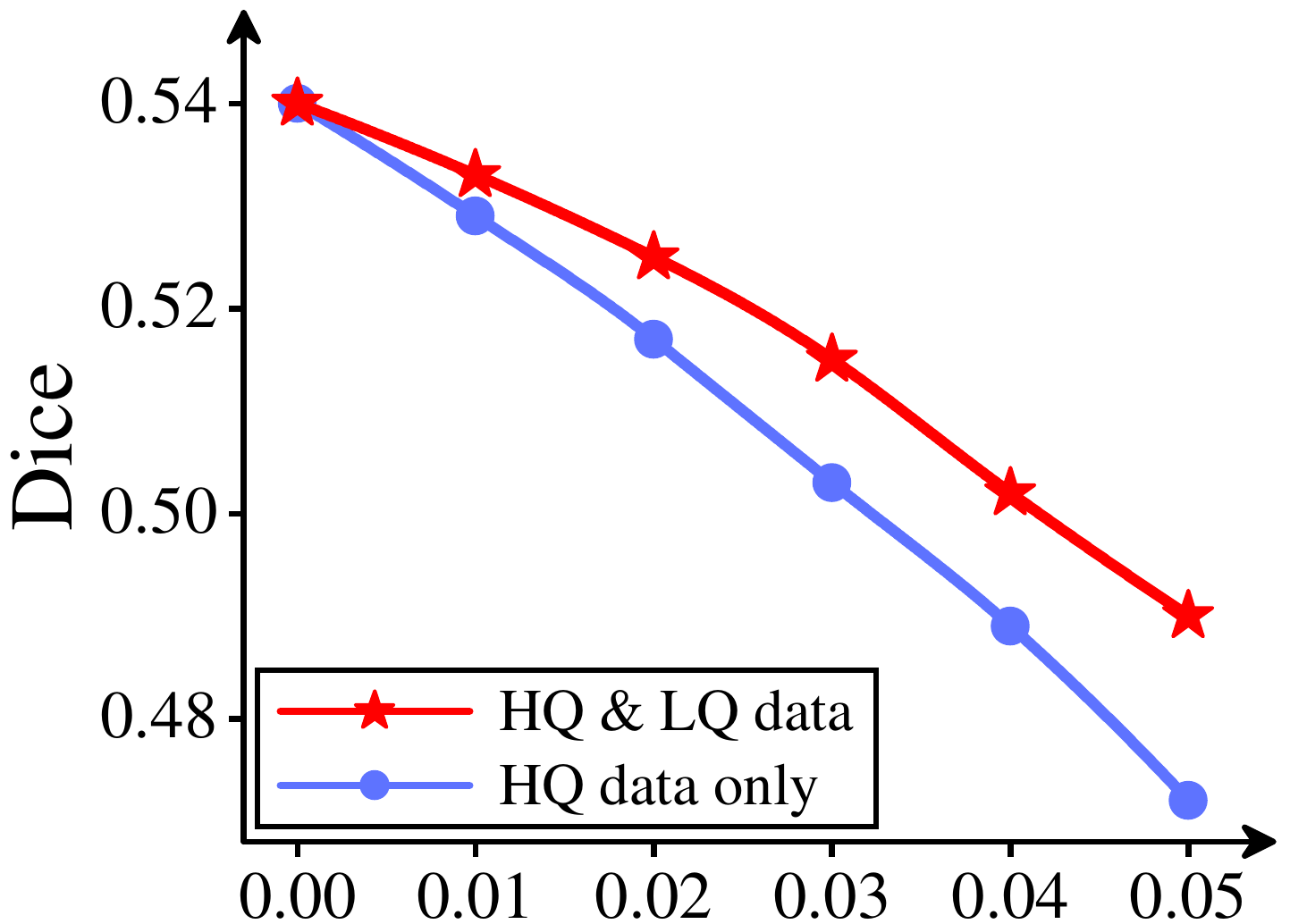}\vspace{-2pt}
		\caption{\footnotesize Poisson noise.}
	\end{subfigure}%\vspace{-1mm}
    \begin{subfigure}{0.163\textwidth}
		\centering
		\includegraphics[width=\textwidth]{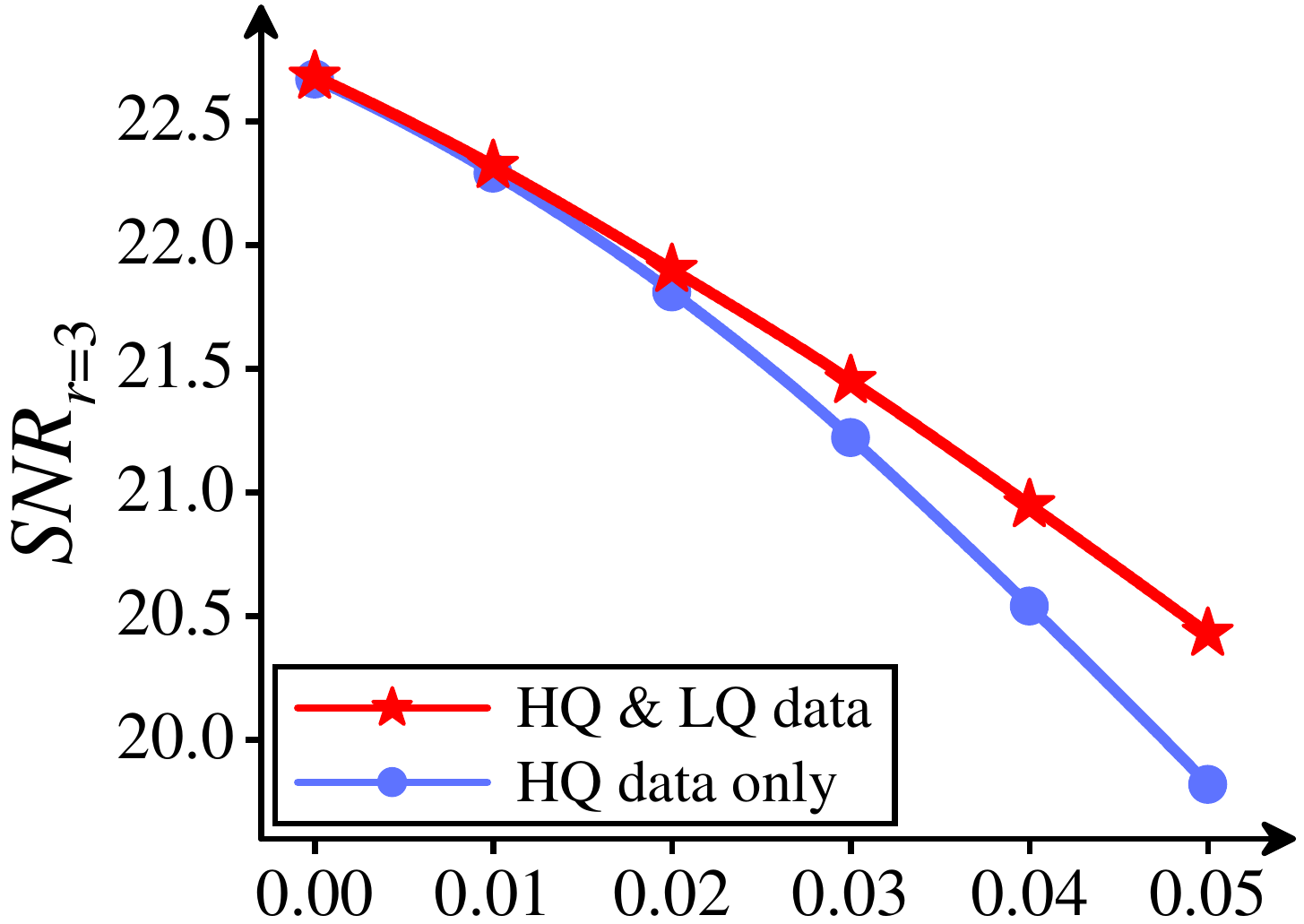}\vspace{-2pt}
		\caption{\footnotesize Salt \& Pepper noise.}
	\end{subfigure}
	\hfill
	\begin{subfigure}{0.163\textwidth}  
		\centering 
		\includegraphics[width=\textwidth]{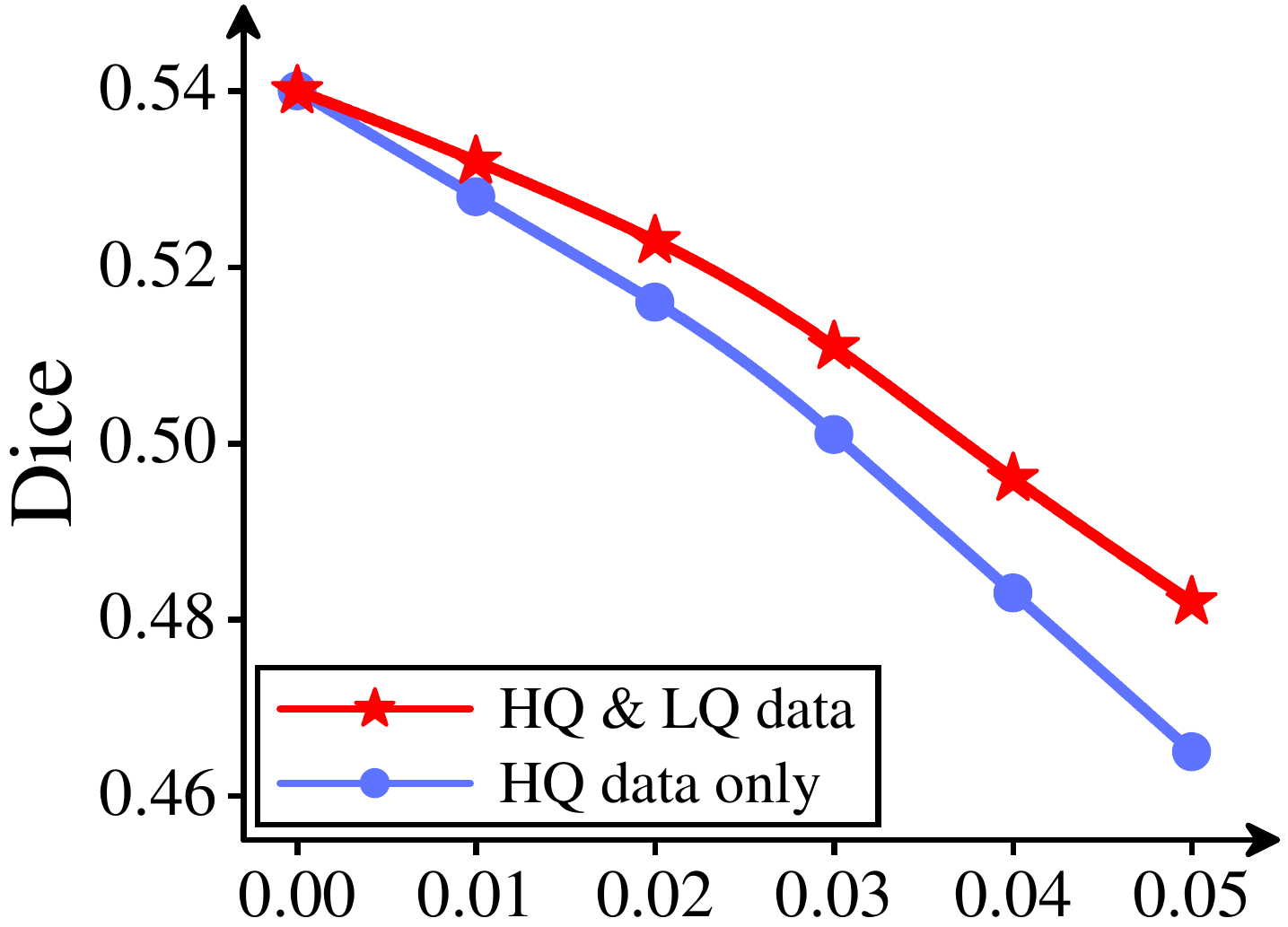}\vspace{-2pt}
		\caption{\footnotesize  Salt \& Pepper noise.}
	\end{subfigure}\vspace{-1mm}
	\caption{Robustness analysis with the HQ cue extraction module trained with different data, where we simulate various kinds of noise for the HQ guidance image in \textit{CCM} dataset and evaluate the enhancement performance with $SNR_{r=3}$ and Dice.}
    \label{fig:RobustAnaHQM}
	\vspace{-4.5mm}
\end{figure*}

\vspace{-2mm}
\subsection{Ablation Study}
%{\color{blue}We conduct extensive ablation studies to verify the effect of our HQ cue extraction module, the variational information normalization module, and the proposed content-aware loss.}

%散开来写，此段删除
%\noindent\textbf{\color{blue}HQ cue extraction module.} {\color{blue}We verify the effect of the HQ cue extraction module and ablate its core components, i.e., the guidance image and the vector-based guidance manner. 

%%散开来写，此段删除 (a) 
%\noindent\textbf{\color{blue}Effect of the HQ cue extraction module.} We prove the effectiveness of the HQ guidance mechanism (Table~\ref{table:EffectHQ}), validate the consistency of the guidance representation (Table~\ref{table:CompactGuidance}), and verify the feasibility to pretrain the HQ cue extraction module with both HQ and LQ data (Fig.~\ref{fig:RobustAnaHQM}). 

\noindent\textbf{Effect of the HQ cue extraction module.} {As shown in Table~\ref{table:EffectHQ}, when directly removing the HQ cue extraction module or replacing the HQ vector with a learnable tensor, the enhancement performance greatly declines, which verifies the effectiveness of our HQ guidance mechanism. 

\noindent\textbf{Consistency of the guidance representation.} Furthermore, to verify the consistency of HQG-Net on the HQ cue extraction, 
%we conduct experiments by incorporating extra constraints to regulate the HQ cue extraction module towards a consistent representation. Specifically, 
we consider regularizing the HQ vector $\textbf{v}_z$ to exhibit minimal variance when the HQ guidance image is altered. To achieve this, we add two constraints, namely $L_{InterVar}$ and $L_{IntraVar}$, to minimize the variance in either different homogeneous HQ guidance images or the same HQ guidance image with different views, which are formulated as:
\begin{equation}
    L_{InterVar}=\left\|Gr(\textbf{v}_{\textbf{z}_1})-Gr(\textbf{v}_{\textbf{z}_2})\right\|_F,
\end{equation}
\begin{equation}
    L_{IntraVar}=\left\|Gr(\textbf{v}_{\textbf{z}^1})-Gr(\textbf{v}_{\textbf{z}^2})\right\|_F,
\end{equation}
where $\textbf{v}_{\textbf{z}_1}$ and $\textbf{v}_{\textbf{z}_2}$ are the vectors of two randomly selected HQ guidance images $\textbf{z}_1$ and $\textbf{z}_2$. $\textbf{v}_{\textbf{z}^1}$ and $\textbf{v}_{\textbf{z}^2}$ are the vectors of HQ images $\textbf{z}^1$ and $\textbf{z}^2$, where $\textbf{z}^2$ is a different view of $\textbf{z}^1$ with the random rotation and flip.
%and $\textbf{z}^2=\mathcal{T}\left(\textbf{z}^1\right)$. $\mathcal{T}\left(\bigcdot\right)$ is the transformation operator for the random rotation and flip. 
As presented in Table~\ref{table:CompactGuidance}, the limited increment from the extra constraints, i.e., $L_{InterVar}$ and $L_{IntraVar}$, indicates that HQG-Net has already learned such consistent representation and thereby extract sufficiently stable HQ cues. This mainly attributes to the explicit regularization of our vector-based HQ guidance mechanism and the implicit constraint of the proposed content-aware loss.

\noindent\textbf{Robustness analysis with the HQ 
 cue extraction module trained with different data.} To verify the feasibility to pretrain the HQ cue extraction module $G$ with both HQ and LQ data, we conduct a robustness analysis for HQG-Net with $G$ pretrained with different data, i.e., ``HQ data only'' and ``HQ \& LQ data'', by simulating different kinds of noise, including Gaussian, Poisson, and Salt \& Pepper noises, for the HQ guidance image. As reported in Fig.~\ref{fig:RobustAnaHQM}, 
%when the guidance image is a real HQ image, there exists little difference between the networks trained with ``HQ data only'' and ``HQ \& LQ data'' on the enhanced results. However, 
$G$ trained with ``HQ data only'' can be inevitably undermined by the fake HQ guidance with perturbations, %, thus failing to extract real HQ cues.
while $G$ trained with ``HQ \& LQ data'' can better resist the noises and extract more stable HQ cues. 
Guided by such HQ cues, HQG-Net can better withstand the complex degradation from the LQ inputs and thereby generate HQ enhanced results. 
%In this case, pretraining the HQ cue extraction module $G$ with both HQ and LQ data facilitates stable extraction of HQ cues and thereby ensures the quality of the enhanced results.

%(b) For the guidance image, we attest to our robustness to the HQ guidance (Table~\ref{table:RobustHQ}) and test our performance under the guidance of different images (Table~\ref{table:RobustAnaGuidance}). 

\noindent\textbf{Robustness to the HQ guidance image.} We first attest to whether the proposed HQG-Net is robust to the HQ guidance image and mainly consider three conditions, i.e., all HQ data from the corresponding test set (200 HQ images), 10 HQ images, and ours (1 fixed image), to randomly select the HQ image to guide the enhancement %of each LQ image 
in the test phase. As shown in Table~\ref{table:RobustHQ}, all three conditions yield similar performance, which evidently verifies our robustness to the HQ guidance image and further indicates that HQG-Net can extract general HQ cues from the HQ guidance images.

\begin{table*}[htbp!]
		\centering
		\caption{Segmentation results on CCM and Colonoscopy datasets, where the best and the second-best are marked in red and blue. In \textit{CCM} dataset, the cooperative training strategy improves the performance by $6.3\%$ (HQG-Net), $2.9\%$ (EnlightenGAN), $3.8\%$ (StillGAN), $2.8\%$ (MASA-SR), and $14.9\%$ (StarGAN V2). In \textit{Colonoscopy} dataset, such a training strategy increases that by $1.3\%$ (HQG-Net), $2.0\%$ (EnlightenGAN), $1.1\%$ (StillGAN), $2.4\%$ (MASA-SR), and $28.2\%$ (StarGAN V2).
		}
		\label{table:segmentation}
		\scalebox{1}{
				% \resizebox{2\columnwidth}{!}{
					\begin{tabular}{l|ccccc|ccccc}
						\toprule[1.5pt]
						\multicolumn{1}{l|}{} 	& \multicolumn{5}{c|}{\textit{CCM}}                                                                                                                                                                                                    & \multicolumn{5}{c}{\textit{Colonoscopy}}                                                                                                                                                                                            \\ \cline{2-11}
						\multicolumn{1}{l|}{\multirow{-2}{*}{Methods}}         & \multicolumn{1}{c|}{\cellcolor{gray!40}AUC$\uparrow$}                          & \multicolumn{1}{c|}{\cellcolor{gray!40}ACC$\uparrow$}                          & \multicolumn{1}{c|}{\cellcolor{gray!40}SEN$\uparrow$}                          & \multicolumn{1}{c|}{\cellcolor{gray!40}Dice$\uparrow$}                         & \cellcolor{gray!40}G\_Mean$\uparrow$                      & \multicolumn{1}{c|}{\cellcolor{gray!40}AUC$\uparrow$}                          & \multicolumn{1}{c|}{\cellcolor{gray!40}ACC$\uparrow$}                          & \multicolumn{1}{c|}{\cellcolor{gray!40}SEN$\uparrow$}                          & \multicolumn{1}{c|}{\cellcolor{gray!40}Dice$\uparrow$}                         & \cellcolor{gray!40}G\_Mean$\uparrow$                      \\ \hline \hline
						Original     & \multicolumn{1}{c|}{0.877}                        & \multicolumn{1}{c|}{{\color[HTML]{00B0F0} \textbf{0.982}}} & \multicolumn{1}{c|}{0.492}                        & \multicolumn{1}{c|}{0.387}                        & 0.674                        & \multicolumn{1}{c|}{0.965}                        & \multicolumn{1}{c|}{{0.978}} & \multicolumn{1}{c|}{0.871}                        & \multicolumn{1}{c|}{0.885}                        & 0.914                        \\ \hline
						CAEFI~\cite{jeelani2019content} & \multicolumn{1}{c|}{0.919}                        & \multicolumn{1}{c|}{0.951}                        & \multicolumn{1}{c|}{0.815}                        & \multicolumn{1}{c|}{0.352}                        & 0.886                        & \multicolumn{1}{c|}{0.959}                        & \multicolumn{1}{c|}{0.974}                        & \multicolumn{1}{c|}{0.841}                        & \multicolumn{1}{c|}{0.856}                        & 0.895                        \\ \hline
						ATA~\cite{zhao2020automated} & \multicolumn{1}{c|}{{0.936}} & \multicolumn{1}{c|}{0.976}                        & \multicolumn{1}{c|}{0.815}                        & \multicolumn{1}{c|}{{0.483}} & {0.869} & \multicolumn{1}{c|}{0.961}                        & \multicolumn{1}{c|}{0.972}                        & \multicolumn{1}{c|}{0.842}                        & \multicolumn{1}{c|}{0.855}                        & 0.891                        \\ \hline
						IDRLP~\cite{ju2021idrlp} & \multicolumn{1}{c|}{0.931}                        & \multicolumn{1}{c|}{0.975}                        & \multicolumn{1}{c|}{0.765}                        & \multicolumn{1}{c|}{0.477}                        & 0.866                        & \multicolumn{1}{c|}{0.764}                        & \multicolumn{1}{c|}{0.914}                        & \multicolumn{1}{c|}{0.407}                        & \multicolumn{1}{c|}{0.407}                        & 0.493                        \\ \hline
						StarGAN V2~\cite{choi2020stargan} & \multicolumn{1}{c|}{0.851}                        & \multicolumn{1}{c|}{0.948}                        & \multicolumn{1}{c|}{0.642}                        & \multicolumn{1}{c|}{0.244}                        & 0.775                        & \multicolumn{1}{c|}{0.792}                        & \multicolumn{1}{c|}{0.926}                        & \multicolumn{1}{c|}{0.411}                        & \multicolumn{1}{c|}{0.459}                        & 0.514                        \\ \hline
                        SSEN~\cite{shim2020robust} & \multicolumn{1}{c|}{0.922}                        & \multicolumn{1}{c|}{0.973}                        & \multicolumn{1}{c|}{0.818}                        & \multicolumn{1}{c|}{0.445}                        & 0.857                        & \multicolumn{1}{c|}{0.967}                        & \multicolumn{1}{c|}{0.972}                        & \multicolumn{1}{c|}{0.854}                        & \multicolumn{1}{c|}{0.877}                        & 0.910                        \\ \hline
                        TTSR~\cite{yang2020learning} & \multicolumn{1}{c|}{0.917}                        & \multicolumn{1}{c|}{0.968}                        & \multicolumn{1}{c|}{0.814}                        & \multicolumn{1}{c|}{0.468}                        & 0.861                        & \multicolumn{1}{c|}{0.965}                        & \multicolumn{1}{c|}{0.964}                        & \multicolumn{1}{c|}{0.847}                        & \multicolumn{1}{c|}{0.894}                        & 0.922                        \\ \hline
                        $C^2$-Matching~\cite{jiang2021robust} & \multicolumn{1}{c|}{0.915}                        & \multicolumn{1}{c|}{0.977}                        & \multicolumn{1}{c|}{0.821}                        & \multicolumn{1}{c|}{0.472}                        & 0.854                        & \multicolumn{1}{c|}{0.973}                        & \multicolumn{1}{c|}{0.968}                        & \multicolumn{1}{c|}{0.851}                        & \multicolumn{1}{c|}{0.872}                        & 0.906                        \\ \hline
                        MASA-SR~\cite{lu2021masa} & \multicolumn{1}{c|}{0.928}                        & \multicolumn{1}{c|}{0.975}                        & \multicolumn{1}{c|}{0.825}                        & \multicolumn{1}{c|}{\color[HTML]{00B0F0}\textbf{0.497}}                        & 0.874                        & \multicolumn{1}{c|}{0.969}                        & \multicolumn{1}{c|}{0.975}                        & \multicolumn{1}{c|}{0.862}                        & \multicolumn{1}{c|}{0.891}                        & 0.918                        \\ \hline
						StillGAN~\cite{ma2021structure} & \multicolumn{1}{c|}{\color[HTML]{00B0F0} \textbf{0.942}}                        & \multicolumn{1}{c|}{0.973}                        & \multicolumn{1}{c|}{\color[HTML]{00B0F0}{\textbf{0.831}}}               & \multicolumn{1}{c|}{0.488}                        & 0.879                        & \multicolumn{1}{c|}{{\color[HTML]{00B0F0} \textbf{0.976}}} & \multicolumn{1}{c|}{\color[HTML]{00B0F0}\textbf{0.979}}                        & \multicolumn{1}{c|}{0.856}                        & \multicolumn{1}{c|}{\color[HTML]{00B0F0}\textbf{0.901}}               & {\color[HTML]{FF0000}\textbf{0.931}}               \\ \hline
						EnlightenGAN~\cite{jiang2021enlightengan} & \multicolumn{1}{c|}{0.929}                        & \multicolumn{1}{c|}{0.969}                        & \multicolumn{1}{c|}{0.771}                        & \multicolumn{1}{c|}{0.457}                        & \color[HTML]{00B0F0} \textbf{0.887}                        & \multicolumn{1}{c|}{0.965}                        & \multicolumn{1}{c|}{0.978}                        & \multicolumn{1}{c|}{{\color[HTML]{00B0F0} \textbf{0.887}}} & \multicolumn{1}{c|}{0.879}                        & 0.907                        \\ \hline
						HQG-Net         & \multicolumn{1}{c|}{\color[HTML]{FF0000}\textbf{0.951}}              & \multicolumn{1}{c|}{\color[HTML]{FF0000}\textbf{0.984}}               & \multicolumn{1}{c|}{{\color[HTML]{FF0000} \textbf{0.837}}} & \multicolumn{1}{c|}{\color[HTML]{FF0000}\textbf{0.536}}            & {\color[HTML]{FF0000}\textbf{0.893}}          & \multicolumn{1}{c|}{\color[HTML]{FF0000}\textbf{0.978}}         & \multicolumn{1}{c|}{\color[HTML]{FF0000}\textbf{0.981}}           & \multicolumn{1}{c|}{\color[HTML]{FF0000}\textbf{0.891}}               & \multicolumn{1}{c|}{{\color[HTML]{FF0000} \textbf{0.906}}} & {\color[HTML]{00B0F0} \textbf{0.930}} \\ \hline \hline
						StarGAN V2+ &\multicolumn{1}{c|}{0.871}&\multicolumn{1}{c|}{0.968}&\multicolumn{1}{c|}{0.685}&\multicolumn{1}{c|}{0.390}&\multicolumn{1}{c|}{0.801}&\multicolumn{1}{c|}{0.836}&\multicolumn{1}{c|}{0.964}&\multicolumn{1}{c|}{0.537}&\multicolumn{1}{c|}{0.663}&0.802 \\ \hline
                        MASA-SR+ &\multicolumn{1}{c|}{0.934}&\multicolumn{1}{c|}{0.976}&\multicolumn{1}{c|}{0.840}&\multicolumn{1}{c|}{0.542}&\multicolumn{1}{c|}{0.883}&\multicolumn{1}{c|}{0.970}&\multicolumn{1}{c|}{0.980}&\multicolumn{1}{c|}{0.888}&\multicolumn{1}{c|}{0.913}&0.930 \\ \hline
                        StillGAN+ &\multicolumn{1}{c|}{0.950}&\multicolumn{1}{c|}{0.971}&\multicolumn{1}{c|}{0.851}&\multicolumn{1}{c|}{0.557}&\multicolumn{1}{c|}{0.886}&\multicolumn{1}{c|}{0.976}&\multicolumn{1}{c|}{0.981}&\multicolumn{1}{c|}{0.878}&\multicolumn{1}{c|}{0.916}&0.941 \\ \hline
                        EnlightenGAN+ &\multicolumn{1}{c|}{0.934}&\multicolumn{1}{c|}{0.972}&\multicolumn{1}{c|}{0.788}&\multicolumn{1}{c|}{0.502}&\multicolumn{1}{c|}{0.891}&\multicolumn{1}{c|}{0.971}&\multicolumn{1}{c|}{0.982}&\multicolumn{1}{c|}{0.896}&\multicolumn{1}{c|}{0.906}&0.922 \\ \hline
						HQG-Net+ &\multicolumn{1}{c|}{0.963} &\multicolumn{1}{c|}{0.991} &\multicolumn{1}{c|}{0.872} &\multicolumn{1}{c|}{0.663} & 0.907 &\multicolumn{1}{c|}{0.982} &\multicolumn{1}{c|}{0.986} &\multicolumn{1}{c|}{0.910} &\multicolumn{1}{c|}{0.924} & 0.943 \\ \bottomrule[1.5pt]
			\end{tabular}}
   % }
			\vspace{-4mm}
		\end{table*}

\noindent\textbf{Using other types of images for guidance}. 
To verify the advantage of using the HQ images from the same dataset for guiding the enhancement process, we replace the HQ images with other types of images, namely, 
(1) HQ heterogeneous medical images from another dataset,
(2) LQ medical data,
(3) natural images. 
Table~\ref{table:RobustAnaGuidance} shows the results. We can see that these alternative choices produce inferior results to using the HQ images from the same dataset as guidance. 

However, as shown in Table~\ref{table:RobustAnaGuidance}, when being guided by HQ heterogeneous medical data, HQG-Net still generates promising enhanced results that exceed
%. These results, although lower in quality than using HQ fundus data as the guidance, still exceed 
most comparison methods, which lie in the similarities of HQ medical images (see Fig.~\ref{fig:HQDefinition}). For example, %both HQ CCM and HQ fundus data possess enhanced texture details, and 
both HQ colonoscopy and HQ fundus data hold uniform illumination. Besides, HQG-Net introduces vector-based HQ guidance to ensure the extraction of valid HQ cues and employs the variational information normalization module to guide the enhancement with the concise and general variational information from the HQ cues. Therefore, even when the relevance of the HQ guidance image to the LQ image is not very tight, HQG-Net can also extract some valuable HQ cues and apply them to improve the enhancement performance.

Additionally, as depicted in Table~\ref{table:RobustAnaGuidance}, even under the guidance of the least relevant and even seriously degraded data, the enhanced performance is gracefully degraded to our Pix2Pix-based version, which demonstrates that introducing the vector-based HQ guidance in a variational manner is a stable and harmless way to improve the enhancement performance.

\begin{figure}[tbh!]
	\centering
	\begin{subfigure}{0.163\textwidth}
		\centering
		\includegraphics[width=\textwidth]{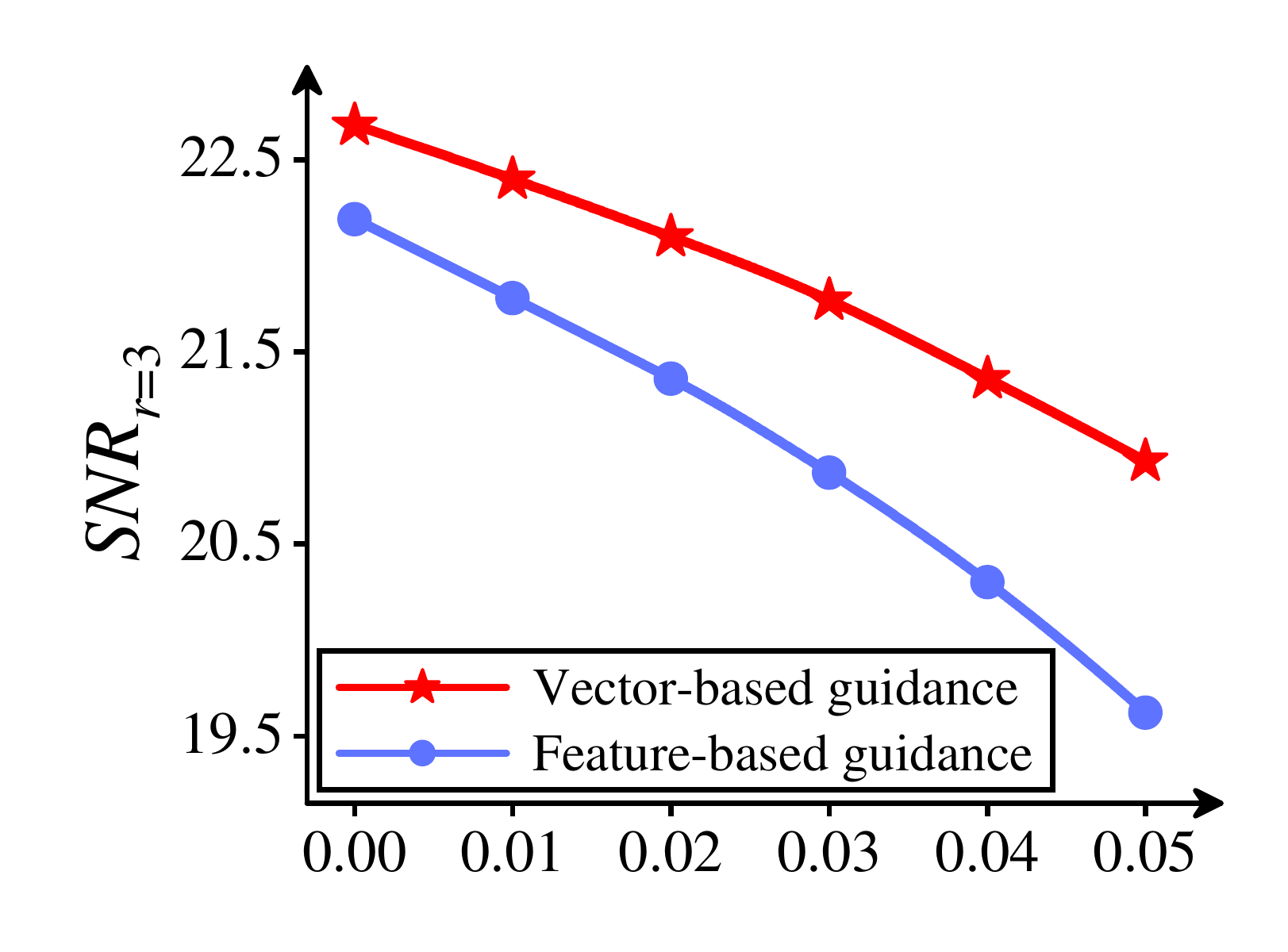}\vspace{-2pt}
		\caption{\footnotesize Gaussian noise.}
	\end{subfigure}\hspace{1cm}
	%\hfill
	\begin{subfigure}{0.163\textwidth}  
		\centering 
		\includegraphics[width=\textwidth]{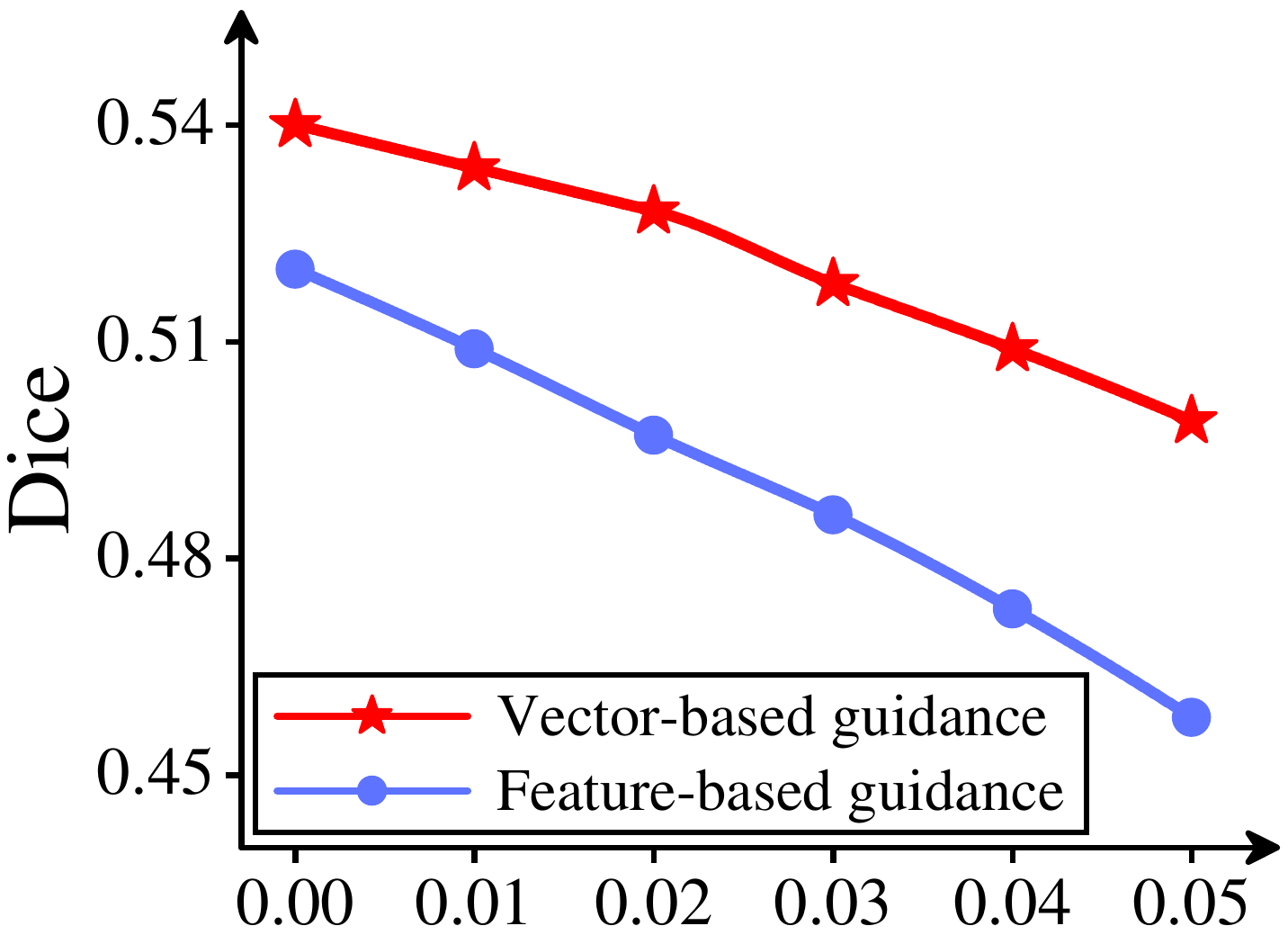}\vspace{-2pt}
		\caption{\footnotesize Gaussian noise.}
	\end{subfigure}\\\vspace{-1mm}
	\caption{Robustness analysis with vector-based guidance (ours) and feature-based guidance, where we simulate Gaussian noise for the HQ guidance image in \textit{CCM} dataset.}
    \label{fig:RobustAnaVector}
	\vspace{-6mm}
\end{figure}

To sum up, the superiority of the enhanced results under the highly relevant HQ guidance, i.e., the HQ homogeneous images, and the robustness of them under the least relevant and seriously degraded guidance, i.e., the LQ natural images, validate that \textit{our guidance-based framework is a meaningful exploration for the UMIE task and can draw the attention of the community to handle the UMIE task with the explicit assistance of reference images}. Besides, considering that it is very easy for a hospital to obtain an HQ image that is highly correlated with the LQ image, e.g., LQ and HQ data acquired by the same instrument with the same set of parameters for the same organs, \textit{our HQG-Net has a high clinical utility}.

%(c) %For the vector-based guidance manner, 
\noindent\textbf{Effect of the vector-based guidance.} We further validate the efficacy of the vector-based guidance with the joint vector $\textbf{v}_c$. In Fig.~\ref{fig:RobustAnaVector}, HQG-Net with vector-based guidance can better enhance LQ images and moderate the slight perturbations from the HQ guidance, which verifies that the vector-based guidance helps to extract valid HQ cues. Besides, as exhibited in Table~\ref{table:EffectAdaLIN}, guided by the joint vector $\textbf{v}_z$ that combines the HQ vector $\textbf{v}_z$ and structure vector $\textbf{v}_c$,
%the joint vector $\textbf{v}_z$ combines the HQ vector $\textbf{v}_z$ and structure vector $\textbf{v}_c$ from the LQ input to provide comprehensive guidance. Hence, as exhibited in Table~\ref{table:EffectAdaLIN}, guided by the joint vector, 
HQG-Net is capable of preserving critical structural information and achieving superior enhancement performance.}

\noindent\textbf{Effect of the variational information normalization module.} {We verify the effect of the variational information normalization module by ablating AdaLIN (Table~\ref{table:EffectAdaLIN}) and comparing AdaLIN with other integration strategies at the feature level (Table~\ref{table:AdaLINVersus}). In Table~\ref{table:EffectAdaLIN}, by canceling the multiscale strategy and ablating AdaLIN, we validate the effect of the multiscale AdaLIN operator.
\begin{figure*}[tbp!]
			\centering
			\includegraphics[width=1\textwidth]{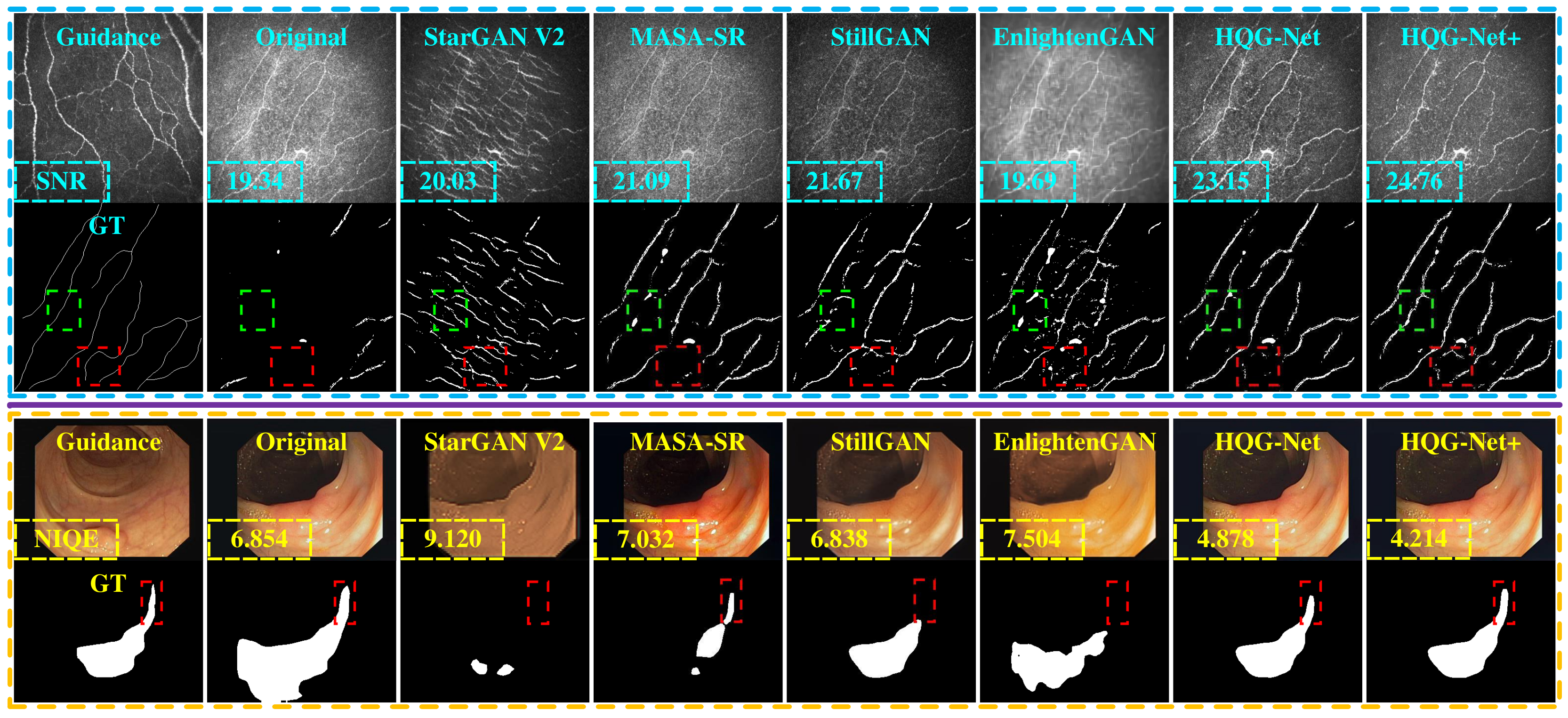}\vspace{-0.3cm}
			\caption{Visual comparison on CCM and Colonoscopy, where guidance images and segmentation results are also presented.}
			\label{fig:CCM_Quanti}
			\vspace{-4.5mm}
		\end{figure*}
  \begin{figure*}[htbh!]
	\centering
	\begin{subfigure}{0.32\textwidth}
		\centering
		\includegraphics[width=\textwidth]{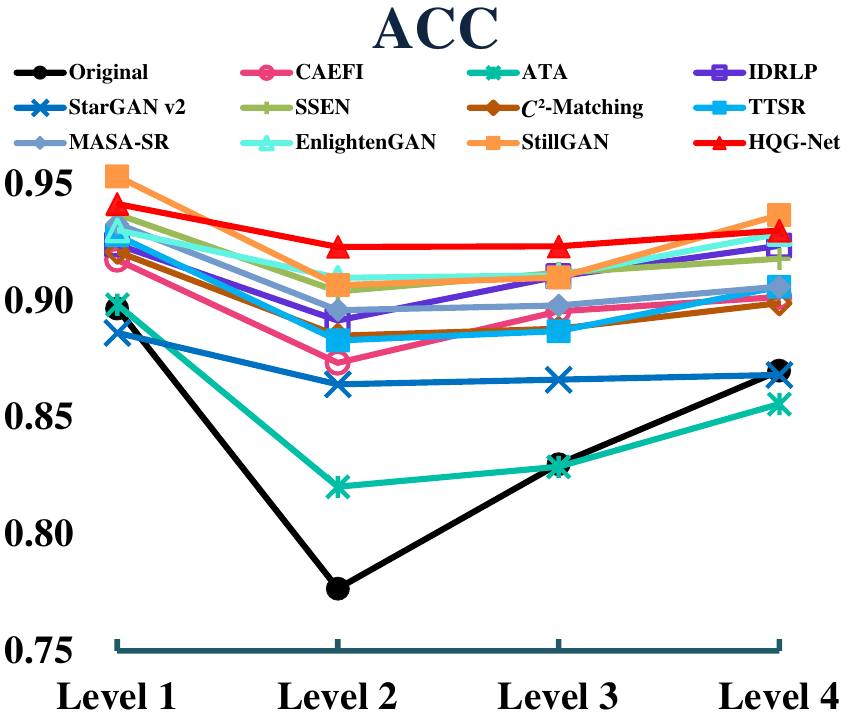}
	\end{subfigure}
	\hfill
	\begin{subfigure}{0.32\textwidth}  
		\centering 
		\includegraphics[width=\textwidth]{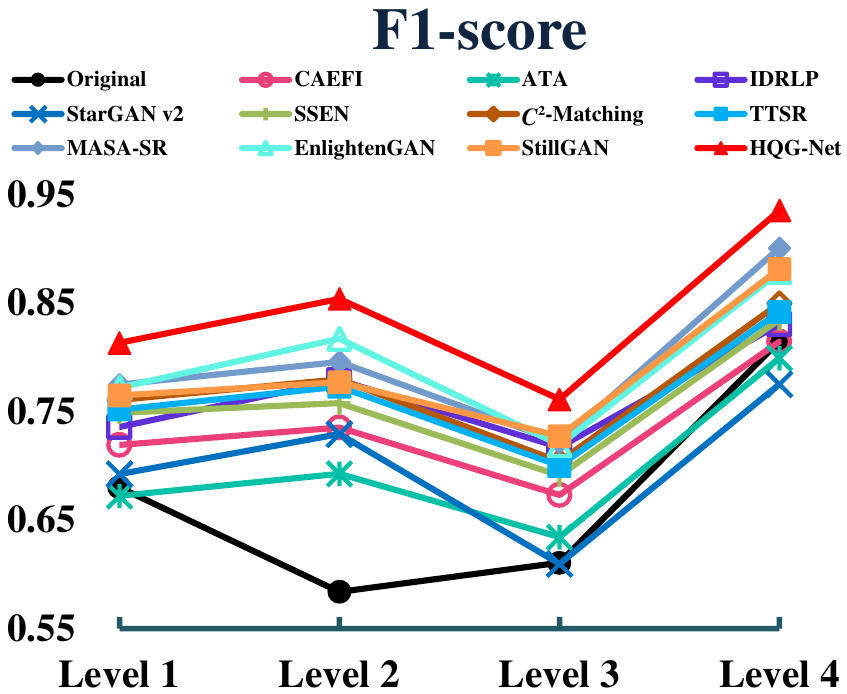}
	\end{subfigure}
	\hfill
	\begin{subfigure}{0.32\textwidth}   
		\centering 
		\includegraphics[width=\textwidth]{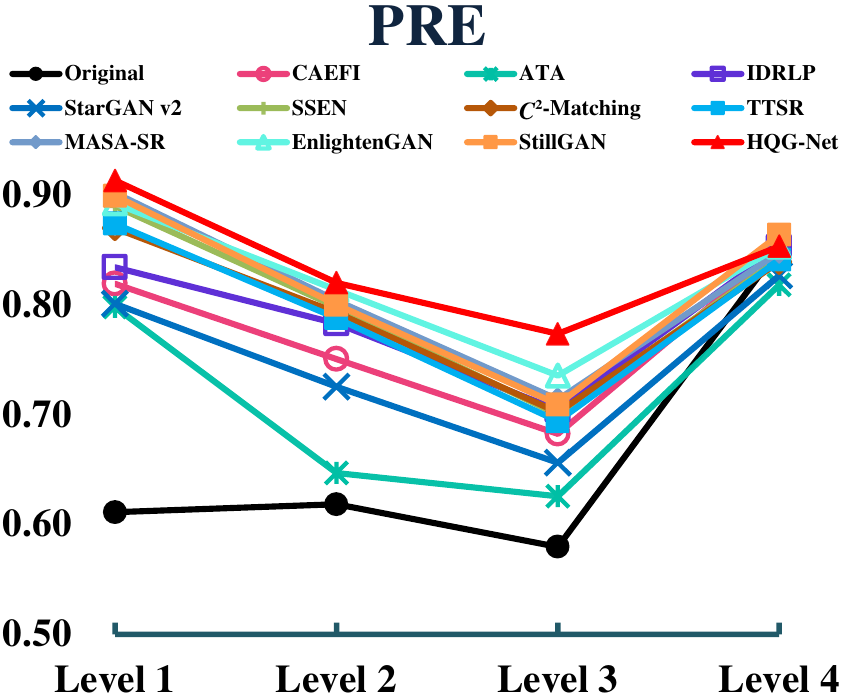}
	\end{subfigure}
	\caption[Nerve Fiber Tortuosity Classification]
	{Nerve fiber tortuosity classification under different methods on CORN-3 dataset with ACC, F1-score, and PRE.} 
	\label{fig:NerveFiberTortuosityClassification}
	\vspace{-2.5mm}
\end{figure*}
%we first cancel the multiscale strategy of the variational instance normalization block in the decoder and conclude that the multiscale strategy benefits the enhancement performance. Then the two ablation studies, i.e., replacing AdaLIN with AdaIN or AdaLN, comprehensively demonstrate that the adaptive AdaLIN operator has a positive compounding effect on HQG-Net. 
Besides, when substituting AdaLIN for other integration strategies, namely feature-level concatenate, add, or multiple, we directly use the combined feature maps rather than the joint vector $\textbf{v}_c$ for feature matching. For fairness, we also provide the results of AdaLIN with the combined features. 
%Besides, to accommodate the multiscale strategy, we adjust the size of the combined feature maps for dimensional alignment. 
Table~\ref{table:AdaLINVersus} validates the superiority of AdaLIN over other information integration strategies, which attributes to the concise and general variational information from AdaLIN.
%This is because our proposed variational information normalization module, i.e., AdaLIN, introduces the concise and general variational information from HQ guidance to guide the enhancement, thus improving the enhancement quality.
}

\begin{table*}[t]
    \centering
    \caption{Generalization of our cooperative training strategy, where Enlight is short for EnlightenGAN and ``+'' denotes optimizing the method under the cooperative training strategy. Increases are marked in red.} \label{table:GeneralizationBLO}\vspace{-1mm}
    \begin{subtable}{.49\textwidth}
		\centering
        \caption{Accuracy of nerve fiber tortuosity classification.}\vspace{-2mm}
		\label{table:GeneralizationBLO-ACC}
    \resizebox{1\columnwidth}{!}{
    \setlength{\tabcolsep}{0.4mm}
    \begin{tabular}{l|cccccc}
    \toprule[1.5pt]
     Tortuosity&\cellcolor{c2!50} HQG-Net & \multicolumn{1}{c|}{\cellcolor{c2!50}HQG-Net+}        &\cellcolor{c2!50} StillGAN &\multicolumn{1}{c|}{\cellcolor{c2!50}StillGAN+}       & \cellcolor{c2!50}Enlight & \multicolumn{1}{c}{\cellcolor{c2!50}Enlight+}       \\ \hline \hline
    Level 1    & 0.941  & \multicolumn{1}{c|}{0.972 ({\color{red}+0.031})} & 0.953    & \multicolumn{1}{c|}{0.978 ({\color{red}+0.025})} & 0.930        & \multicolumn{1}{c}{0.961 ({\color{red}+0.031})} \\ \hline
    Level 2    & 0.923  & \multicolumn{1}{c|}{0.931 ({\color{red}+0.008})} & 0.906    & \multicolumn{1}{c|}{0.922 ({\color{red}+0.016})} & 0.909        & \multicolumn{1}{c}{0.920 ({\color{red}+0.011})} \\ \hline
    Level 3    & 0.923  & \multicolumn{1}{c|}{0.947 ({\color{red}+0.024})} & 0.910    & \multicolumn{1}{c|}{0.931 ({\color{red}+0.021})} & 0.911        & 0.926 ({\color{red}+0.015}) \\ \hline
    Level 4    & 0.930  & \multicolumn{1}{c|}{0.952 ({\color{red}+0.022})} & 0.936    & \multicolumn{1}{c|}{0.949 ({\color{red}+0.013})} & 0.929        & 0.946 ({\color{red}+0.017}) \\ \bottomrule[1.5pt]
    \end{tabular}}
    \end{subtable}
    \begin{subtable}{.49\textwidth}
		\centering
        \caption{F1-score of nerve fiber tortuosity classification.}\vspace{-2mm}
		\label{table:GeneralizationBLO-F1}
    \resizebox{1\columnwidth}{!}{
    \setlength{\tabcolsep}{0.4mm}
    \begin{tabular}{l|cccccc}
    \toprule[1.5pt]
     Tortuosity&\cellcolor{c2!50} HQG-Net & \multicolumn{1}{c|}{\cellcolor{c2!50}HQG-Net+}        &\cellcolor{c2!50} StillGAN &\multicolumn{1}{c|}{\cellcolor{c2!50}StillGAN+}       & \cellcolor{c2!50}Enlight & \multicolumn{1}{c}{\cellcolor{c2!50}Enlight+}       \\ \hline \hline
    Level 1    & 0.813  & \multicolumn{1}{c|}{0.835 ({\color{red}+0.022})} & 0.765    & \multicolumn{1}{c|}{0.796 ({\color{red}+0.031})} & 0.772        & \multicolumn{1}{c}{0.785 ({\color{red}+0.013})} \\ \hline
    Level 2    & 0.854  & \multicolumn{1}{c|}{0.876 ({\color{red}+0.022})} & 0.777    & \multicolumn{1}{c|}{0.813 ({\color{red}+0.036})} & 0.817        & \multicolumn{1}{c}{0.858 ({\color{red}+0.041})} \\ \hline
    Level 3    & 0.761  & \multicolumn{1}{c|}{0.811 ({\color{red}+0.050})} & 0.727    & \multicolumn{1}{c|}{0.771 ({\color{red}+0.044})} & 0.719        & 0.764 ({\color{red}+0.045}) \\ \hline
    Level 4    & 0.935  & \multicolumn{1}{c|}{0.942 ({\color{red}+0.007})} & 0.882    & \multicolumn{1}{c|}{0.906 ({\color{red}+0.024})} & 0.879        & 0.895 ({\color{red}+0.016}) \\ \bottomrule[1.5pt]
    \end{tabular}}
    \end{subtable}\vspace{-2mm}
\end{table*}
\begin{table*}[t]
    \centering
    \caption{Medical image reclassification of HQG-Net and HQG-Net+ on the enhanced LQ data and re-enhanced HQ data in \textit{CCM} and \textit{Colonoscopy} datasets. 
    The table shows the count of images that can be classified as HQ data from the experts' view.}\label{table:Reclass} \vspace{-1mm}
    \begin{subtable}{.49\textwidth}
		\centering
        \caption{Image reclassification on the enhanced LQ data.}\vspace{-2mm}
		\label{table:ReclassLQ}
    \resizebox{\columnwidth}{!}{
    \setlength{\tabcolsep}{1mm}
    \begin{tabular}{l|c|c|c|c|c|c}
    \toprule[1.5pt]
\multirow{2}{*}{Experts} & \multicolumn{3}{c|}{\textit{CCM} (60 images)} & \multicolumn{3}{c}{\textit{Colonoscopy} (70 images)} \\ \cline{2-7}
&\multicolumn{1}{c|}{\cellcolor{c2!50}Original}& \multicolumn{1}{c|}{\cellcolor{c2!50}HQG-Net}& \cellcolor{c2!50}HQG-Net+&\multicolumn{1}{c|}{\cellcolor{c2!50}Original} & \multicolumn{1}{c|}{\cellcolor{c2!50}HQG-Net}& \cellcolor{c2!50}HQG-Net+               \\ \hline \hline
Clinician 1        &0      & \multicolumn{1}{c|}{43}             & 47 ({\color{red}+9.3\%})     &0   & \multicolumn{1}{c|}{55}                 & 59 ({\color{red}+7.3\%})            \\ \hline
Clinician 2        &0      & \multicolumn{1}{c|}{46}             & 51 ({\color{red}+10.9\%})    &0   & \multicolumn{1}{c|}{58}                 & 61 ({\color{red}+5.2\%})  \\ \bottomrule[1.5pt]       
\end{tabular}}
    \end{subtable}
    \begin{subtable}{.49\textwidth}
		\centering
        \caption{Image reclassification on the re-enhanced HQ data.}\vspace{-2mm}
		\label{table:ReclassHQ}
    \resizebox{\columnwidth}{!}{
    \setlength{\tabcolsep}{1mm}
    \begin{tabular}{l|c|c|c|c|c|c}
    \toprule[1.5pt]
\multirow{2}{*}{Experts} & \multicolumn{3}{c|}{\textit{CCM} (200 images)} & \multicolumn{3}{c}{\textit{Colonoscopy} (200 images)} \\ \cline{2-7}
&\multicolumn{1}{c|}{\cellcolor{c2!50}Original} & \multicolumn{1}{c|}{\cellcolor{c2!50}HQG-Net}&\cellcolor{c2!50}HQG-Net+ &\multicolumn{1}{c|}{\cellcolor{c2!50}Original}& \multicolumn{1}{c|}{\cellcolor{c2!50}HQG-Net}& \cellcolor{c2!50}HQG-Net+               \\ \hline \hline
Clinician 1              & \multicolumn{1}{c|}{200}& \multicolumn{1}{c|}{200}              & 200               & \multicolumn{1}{c|}{200} &  \multicolumn{1}{c|}{200}                & 200 \\ \hline
Clinician 2              & \multicolumn{1}{c|}{200}& \multicolumn{1}{c|}{200}               & 200               & \multicolumn{1}{c|}{200} &  \multicolumn{1}{c|}{200}                & 200  \\ \bottomrule[1.5pt] 
\end{tabular}}%\vspace{-2mm}
    \end{subtable}\vspace{-1mm}
\end{table*}	
\begin{table*}[t]
    \centering
    \caption{Disease diagnosis of original LQ data and those enhanced by our HQG-Net on the newly collected \textit{DiaRet} dataset.}\label{table:DiseaseDia} \vspace{-1mm}
    \begin{subtable}{.245\textwidth}
		\centering
        \caption{ACC of disease diagnosis.}\vspace{-2mm}
		\label{table:DiseaseDiaACC}
    \resizebox{1\columnwidth}{!}{
    \setlength{\tabcolsep}{0.65mm}
    \begin{tabular}{l|c|c}\toprule[1.2pt]
Experts         & \cellcolor{c2!50}Original & \cellcolor{c2!50}HQG-Net \\ \hline \hline
Clinician 1 & 0.745    & 0.825 ({\color{red}+10.7\%}) \\ \hline
Clinician 2 & 0.735    & 0.830 ({\color{red}+12.9\%}) \\ \bottomrule[1.2pt]
\end{tabular}}\vspace{-3mm}
    \end{subtable}
    \begin{subtable}{.245\textwidth}
		\centering
        \caption{SEN of disease diagnosis.}\vspace{-2mm}
		\label{table:DiseaseDiaSEN}
    \resizebox{1\columnwidth}{!}{
    \setlength{\tabcolsep}{0.65mm}
    \begin{tabular}{l|c|c}\toprule[1.2pt]
Experts         & \cellcolor{c2!50}Original & \cellcolor{c2!50}HQG-Net \\ \hline \hline
Clinician 1 & 0.810     & 0.840 ({\color{red}+3.7\%}) \\ \hline
Clinician 2 & 0.820     & 0.845 ({\color{red}+3.0\%}) \\ \bottomrule[1.2pt]
\end{tabular}}\vspace{-3mm}
    \end{subtable}
 \begin{subtable}{.245\textwidth}
		\centering
        \caption{F1-score of disease diagnosis.}\vspace{-2mm}
		\label{table:DiseaseDiaF1}
    \resizebox{1\columnwidth}{!}{
    \setlength{\tabcolsep}{0.65mm}
    \begin{tabular}{l|c|c}\toprule[1.2pt]
Experts         & \cellcolor{c2!50}Original & \cellcolor{c2!50}HQG-Net \\ \hline \hline
Clinician 1 & 0.755    & 0.820 ({\color{red}+8.6\%}) \\ \hline
Clinician 2 & 0.761    & 0.836 ({\color{red}+9.7\%}) \\ \bottomrule[1.2pt]
\end{tabular}}\vspace{-3mm}
    \end{subtable}
 \begin{subtable}{.245\textwidth}
		\centering
        \caption{PRE of disease diagnosis.}\vspace{-2mm}
		\label{table:DiseaseDiaPRE}
    \resizebox{1\columnwidth}{!}{
    \setlength{\tabcolsep}{0.65mm}
    \begin{tabular}{l|c|c}\toprule[1.2pt]
Experts         & \cellcolor{c2!50}Original & \cellcolor{c2!50}HQG-Net \\ \hline \hline
Clinician 1 & 0.727    & 0.815 ({\color{red}+12.1\%}) \\ \hline
Clinician 2 & 0.719    & 0.844 ({\color{red}+17.4\%}) \\ \bottomrule[1.2pt]
\end{tabular}}\vspace{-3mm}
    \end{subtable} \vspace{-2mm}
\end{table*}	

\noindent\textbf{Effect of the content-aware loss.} {We further ablate the content-aware loss $L_{CA}$ (Table~\ref{table:AblationCAloss}) and evaluate the optimal combination in $L_{Haar}$ (Table~\ref{table:CombinationHF}). In Table~\ref{table:AblationCAloss}, we compare several combinations of loss functions, where $L_{s}$ is the structure loss in StillGAN~\cite{ma2021structure}. Table \ref{table:AblationCAloss} attests to the superiority of our content-aware loss. Besides, as shown in Table~\ref{table:CombinationHF}, we try the different combinations of high-frequency components in $L_{Haar}$ and find that combining all the high-frequency sub-bands results in optimal enhancement performance.}

 \subsection{Downstream Tasks}
% We conduct two experiments, i.e., medical image segmentation and nerve fiber tortuosity classification tasks, to evaluate the usefulness of the enhanced images for downstream tasks.
 
\noindent\textbf{Medical image segmentation task.} Thanks to the paired segmented label for the \textit{CCM} and \textit{Colonoscopy} datasets, we further test the segmentation results with the corresponding segmentors, i.e., CS-Net~\cite{mou2019cs} and PNS-Net~\cite{ji2021progressively} with their pre-trained models. Five commonly used metrics are applied for assessment, i.e., area under the ROC curve (AUC), accuracy (ACC), sensitivity (SEN), Dice coefficient (Dice), and G-mean score (G-Mean), where a higher value indicates a better performance. As shown in Table \ref{table:segmentation}, the proposed HQG-Net achieves nine best values and one second-best value in the ten scores, demonstrating that the image enhanced by our method can best facilitate the downstream segmentation task.  As shown in Fig. \ref{fig:CCM_Quanti}, the segmented results of HQG-Net are more accurate and continuous, which further verifies the superiority of HQG-Net for the highlight of the geometric structure, whether the foreground is filamentous or blocky. 
{Moreover, in Table \ref{table:segmentation}, compared with networks optimized solely with the enhancement loss, the corresponding network optimized by our cooperative training strategy achieves better performance, indicating the potential of the cooperative training strategy to generate enhancement results that are segmentation-friendly.}
 
\noindent\textbf{Nerve fiber tortuosity classification task.} Previous studies indicate that the tortuosity level grading in CCM is highly relevant to some diseases~\cite{zhou2021memorizing,he2019image}, e.g., diabetic neuropathy and hypertensive retinopathy. Hence, we conduct an experiment on nerve fiber tortuosity with a SOTA classifier, Deepgrading~\cite{mou2022deepgrading}, to explore how the enhancement task contributes to disease-related classifications on \textit{CORN-3} dataset~\cite{ma2021structure}, which contains about 300 images that are divided into four levels according to the tortuosity level. Three metrics are used to estimate the performance, i.e., ACC, F1-score, and precision (PRE), where a higher value means better performance. In Fig. \ref{fig:NerveFiberTortuosityClassification}, the proposed HQG-Net achieves ten best and two second-best values among the twelve metrics, which shows that the proposed HQG-Net can improve the identification rate of nerve fiber tortuosity and further demonstrate the clinical value of HQG-Net. {In addition, as shown in Table~\ref{table:GeneralizationBLO}, the cooperative training strategy improves the accuracy of nerve fiber tortuosity classification by $2.3\%$ (HQG-Net), $2.0\%$ (StillGAN), $2.1\%$ (EnlightenGAN), and increases the F1-score of that by $3.2\%$ (HQG-Net), $4.4\%$ (StillGAN), $3.7\%$ (EnlightenGAN). These increments provide strong evidence of the superiority of the cooperative training strategy.}

\vspace{-2mm}
\subsection{Clinical Decision-Making}
%{We consult with our collaborative experts to assess whether the LQ images enhanced by HQG-Net can facilitate clinical decision-making and conduct two experiments, namely medical image reclassification and disease diagnosis tasks.}

\noindent\textbf{Medical image reclassification task.} {In the image reclassification task, we invite two collaborative clinicians, who have participated in the production of our private datasets, to reclassify the LQ images (60 in \textit{CCM} and 70 in \textit{Colonoscopy}) and the HQ images (200 in \textit{CCM} and 200 in \textit{Colonoscopy}) enhanced by HQG-Net and HQG-Net+, where we randomly select homogeneous HQ guidance in the enhancement. To avoid bias from the experts, we leave out that these images have already been enhanced. As shown in Table~\ref{table:ReclassLQ}, both HQG-Net and HQG-Net+ achieve promising performance in this reclassification task, which verifies that HQG-Net and HQG-Net+ have been successful in improving the quality of most LQ images from a clinical perspective. While in Table~\ref{table:ReclassHQ}, none of the enhanced HQ images are identified as the LQ ones, thereby confirming that both HQG-Net and HQG-Net+ do not weaken the quality of the enhanced images.}

\noindent\textbf{Disease diagnosis task.} {In the disease diagnosis task, we create a new dataset, \textit{DiaRet}, comprised of 300 LQ fundus images, obtained from 150 healthy eyes and 150 eyes affected by diabetic retinopathy. These images were selected from the ``usable'' grade of the Eye-Quality~\cite{fu2019evaluation} with their labels, i.e., with or without diabetic retinopathy. We invite two ophthalmologists to diagnose diabetic retinopathy from the given images, i.e., the original data and the data enhanced by our HQG-Net. To minimize subjective factors, the ophthalmologists are invited to perform the diagnostic task based on the original data first and to review the enhanced images three days later. The diagnostic results from the two ophthalmologists are presented in Table~\ref{table:DiseaseDia}, where HQG-Net improves the diagnosis results by $11.8\%$ (ACC), $3.4\%$ (SEN), $9.2\%$ (F1-score), and $14.8\%$ (PRE). Such results firmly demonstrate that HQG-Net is a promising method for facilitating clinical decision-making.}

\section{Conclusion}
In this paper, we propose a GAN-based network for unpaired medical image enhancement with the explicit guidance of unpaired HQ images (HQG-Net), which is the first work to model the medical image enhancement task under the joint distribution between both the LQ and HQ domains. The joint distribution-based framework implicitly contains more comprehensive and unbiased information than those methods, which only rely on the LQ domain. Therefore, HQG-Net can generate enhanced images with better visual fidelity. For high image contrast and enhanced textural details, the content-aware loss function is proposed with both wavelet-based pixel-level and feature-level consistency constraints. We further propose a bi-level formulation to jointly optimize the UMIE framework with the downstream tasks. Compared with the state-of-the-art techniques, the proposed HQG-Net achieves the best performance in the three distinct modalities both in the medical enhancement problem and the subsequent applicability validation tasks. $\textit{Fundus}$ and $\textit{Colonoscopy}$ datasets with enhanced quality labels will be released to the public with the source code in the future for community research.

\section{ACKNOWLEDGMENT} 
The authors would like to express their sincere appreciation to the anonymous reviewers for their insightful comments,
which greatly improved the quality of this paper.

\bibliographystyle{IEEEtran}
\bibliography{VG-GAN}
\end{document}